\documentclass{article}

\usepackage{arxiv}
\usepackage{subcaption}
\usepackage{graphicx}
\usepackage{tikz}
\usepackage{overpic}
\usepackage{rotating}
\usepackage{xfrac}
\usepackage{multicol}
\usepackage{multirow}
\usepackage{placeins}
\usepackage{amssymb}
\usepackage{booktabs}
\usepackage{amsmath}
\usepackage{enumitem}
\usepackage{adjustbox}
\usepackage{lscape}
\usepackage[colorlinks=true,linkcolor=blue,citecolor=blue]{hyperref}
\usetikzlibrary{shapes, arrows.meta, positioning}

\usepackage{bm}

\title{Global sensitivity analysis with multifidelity Monte Carlo and polynomial chaos expansion for carotid artery haemodynamics}

\author{
Friederike Sch\"afer$^{1}$, \quad Daniele E. Schiavazzi$^{2}$, \quad Leif Rune Hellevik$^{1}$, \quad Jacob Sturdy$^{1}$\\
\\
$^{1}$Division of Biomechanics\\ Norwegian University of Science and Technology (NTNU)\\ Norway\\ 
\\
$^{2}$Department of Applied Mathematics and Computational Statistics \\ University of Notre Dame\\ United States
}

\begin{document}
\maketitle

\begin{abstract}
Computational models of the cardiovascular system are increasingly used for the diagnosis, treatment, and prevention of cardiovascular disease. 
Before being used for translational applications, the predictive abilities of these models need to be thoroughly demonstrated through verification, validation, and uncertainty quantification. 
When results depend on multiple uncertain inputs, sensitivity analysis is typically the first step required to separate relevant from unimportant inputs, and is key to determine an initial reduction on the problem dimensionality that will significantly affect the cost of all downstream analysis tasks.
For computationally expensive models with numerous uncertain inputs, sample-based sensitivity analysis may become impractical due to the substantial number of model evaluations it typically necessitates.
To overcome this limitation, we consider recently proposed Multifidelity Monte Carlo estimators for Sobol' sensitivity indices, and demonstrate their applicability to an idealized model of the common carotid artery. 
Variance reduction is achieved combining a small number of three-dimensional fluid-structure interaction simulations with affordable one- and zero-dimensional reduced order models.
These multifidelity Monte Carlo estimators are compared with traditional Monte Carlo and polynomial chaos expansion estimates.
Specifically, we show consistent sensitivity ranks for both bi- (1D/0D) and tri-fidelity (3D/1D/0D) estimators, and superior variance reduction compared to traditional single-fidelity Monte Carlo estimators for the same computational budget.
As the computational burden of Monte Carlo estimators for Sobol' indices is significantly affected by the problem dimensionality, polynomial chaos expansion is found to have lower computational cost for idealized models with smooth stochastic response.
\end{abstract}


\maketitle

\section{Introduction}

Cardiovascular disease (CVD) is the leading cause of death worldwide~\cite{smith_our_2012}. 
It is predicted that the number of deaths due to CVD will increase in the next years due to an increase in the world population, increase in life expectancy, and falling birth rates. 
Thus, there is a need for improved prevention, diagnosis, and treatment of CVD~\cite{laslett_worldwide_2012}. 
In this context, computational research focused on new approaches to combine clinical measurements with physics-based mechanistic models of the cardiovascular system to aid clinicians in decisions regarding diagnosis, treatment and prevention.
These efforts have led to innovative approaches in clinical data acquisition, image analysis and segmentation, boundary condition tuning, and the estimation of in-vivo material parameters.
Building upon this foundation, cardiovascular models are formulated  through the discretization of the Navier-Stokes equations, the structural dynamics equations, and their coupling \cite{QuarteroniAlfio2004MMaN}. These models are commonly solved using high-performance computing on distributed architectures.

A rich literature exists for applications of cardiovascular models, including studies on the (patho-)physiology of the human body~\cite{taylor_finite_1998, liang_closed-loop_2005, olufsen_dynamics_2002}, non-invasive diagnosis of coronary artery disease~\cite{fossan_uncertainty_2018,zarins_computed_2013}, vascular access in hemodialysis patients~\cite{caroli_validation_2013}, analysis of alternative configurations for arteriovenous fistula~\cite{bode_patient-specific_2012}, and cardiovascular bypass surgery~\cite{steele_vivo_2001}.
Numerical models were also used to investigate and optimize the performance of medical devices, such as blood pumps~\cite{fiusco_blood_2022}, drainage cannulae~\cite{fiusco_numerical_2023}, and coronary stents~\cite{migliavacca_expansion_2007}. 
Others include clinical training~\cite{alderliesten_simulation_2004} or inference of clinically relevant parameters such as arterial compliance~\cite{stergiopulos_simple_1994} and total arterial inertance~\cite{stergiopulos_total_1999}, that are not directly measurable.

All these applications rely on clinically acquired data, such as pressures, flow rates, and the characterization of patient-specific anatomy, which are inherently uncertain.
Given that boundary conditions cannot be directly measured in patients, their estimation introduces an additional layer of uncertainty into the model predictions.
Incorporating population averages as model parameters, introduces yet another source of uncertainty.
To meet regulatory clearance requirements, it is essential to quantify the impact of these uncertainties and variations on model predictions through robust approaches in verification, validation, and uncertainty quantification (UQ)~\cite{anderson_verification_2007,food_and_drug_administration_assessing_2021}.  

Uncertainty quantification is the process of systematically characterizing and reducing uncertainties in computational models, including the propagation of uncertainties in model parameters and boundary conditions through the model, to quantify their impact on output uncertainty. SA associates output variability to specific uncertain inputs.
Joint UQ and SA analysis has been shown to improve model robustness~\cite{anderson_verification_2007}, to reduce input variance through input prioritization, and to inform on unimportant parameters which can be kept constant by setting them equal to population averages~\cite{saltelli_global_2008}.
Established and widely used SA methods include Monte Carlo (MC) simulation (see, e.g.,~\cite{mcbook}) and polynomial chaos (PC) expansion~\cite{xiu2002wiener,sudret2008global}.
Both these approaches were applied to cardiovascular models for estimating total arterial compliance and fractional flow reserve (FFR)~\cite{eck_guide_2016,fossan_uncertainty_2018,tanade_global_2021}. 
Other studies have focused on non-linear material models for one-dimensional wave propagation networks~\cite{eck_effects_2017}, personalized cerebrovascular models for elderly patients with dementia~\cite{zhang_personalized_2021}, studies of the carotid bifurcation~\cite{gul_parametric_2015}, and in arterovenous fistula surgery~\cite{huberts_sensitivity_2013}.

However, use of UQ and SA is severely limited for computationally expensive high-fidelity models, that are essential to resolve \emph{local} quantities of interest (QoIs) such as complex flow patterns (e.g., recirculation) or wall shear stress.
Multifidelity estimators were therefore proposed, leading to a reduced variance with respect to single-fidelity MC estimators for the same computational cost~\cite{peherstorfer_optimal_2016}. 
Cardiovascular applications of multifidelity estimators include quantification of FFR~\cite{yin_one-dimensional_2019}, fast UQ of activation sequences in patient-specific cardiac electrophysiology~\cite{quaglino_fast_2018}, and coronary artery disease~\cite{seo_multifidelity_2020}. 
Combined multilevel and multifidelity estimators for  aorto-femoral and coronary artery models are investigated in~\cite{fleeter_multilevel_2020}.

Approaches for multifidelity SA have so far not been studied as extensively as UQ methods. Qian et al.~\cite{qian_multifidelity_2018} proposed a multifidelity estimator for the main and total Sobol' indices, demonstrating its advantages on the Ishigami function and a 2D hydrogen combustion model.
This method has also been applied to nonlinear Schroedinger and Burgers equations~\cite{du_uncertainty_2022}, to determine particle distributions in turbulent pipe flow~\cite{yao_multi-fidelity_2022}, and to analyze thermal distortions in the James Webb Space Telescope~\cite{cataldo_multifidelity_2022}. 
In the medical field, multifidelity SA has been used in cardiac electrophysiology~\cite{quaglino_high-dimensional_2019,pagani_enabling_2021}, but not for the analysis of vacular flow with fluid-structure interaction (FSI).

In this paper, we compute multifidelity Sobol' indices following the approach proposed in Qian et al.~\cite{qian_multifidelity_2018} for an idealized three-dimensional model of the common carotid artery (CCA) with Arbitrary Lagrangian-Eulerian FSI coupling.
Specifically, our analysis focuses on quantifying the effect of uncertainty in the vessel diameter, wall thickness, and elastic modulus on the systolic pressure~$P_{sys}$, pulse pressure PP, and maximal change in vessel radius~$\Delta r_{\text{max}}$.
We characterize variance reduction for bi- and tri-fidelity estimators, and compare the Sobol' indices with results from MC and PC expansion. The implementation of the MFMC SA framework for the CCA can be found at \url{https://gitlab.com/fe_ls/mulifidelity-mc}. 
The novelty in this work lies in the application of the proposed multifidelity framework to high-fidelity haemodynamic problems with FSI, comparison between estimators formulated using two and three fidelities, the use of perturbed low-fidelity models to overcome issues with the computation of model allocations, and a comparison between sensitivity estimates with MC and PC.

The paper is structured as follows. Section~\ref{sec:methods} offers an overview of global sensitivity analysis, introduces MFMC estimators for Sobol' indices, and presents formulations for the physics-based solvers utilized at each fidelity. 
In Section~\ref{sec:bc_inputs_qoi}, we define the boundary conditions, uncertain inputs, and the selected QoIs for the study. 
Section~\ref{sec:results} focuses on the results from the analysis, including an initial mesh convergence study, statistics and optimal allocations from pilot runs, and sensitivity indices from both bi- and tri-fidelity estimators. 
Section~\ref{sec:comp_cost} compares the computational costs of MFMC, MC, and PC. A discussion and some conclusions are outlined in Sections~\ref{sec:discussion}.

\section{Methods}\label{sec:methods}

\subsection{Global sensitivity analysis}

Global sensitivity analysis (SA) quantifies the importance of the inputs of a computational model through a decomposition of the output variance. 
This class of methods has several advantages including the ability to account for the full range of variation of each uncertain input, their interaction~\cite{saltelli_global_2008}, and is able to capture non-additive, non-monotonic, and non-linear input-output dependencies~\cite{eck_guide_2016}. 
Under the assumption of independent random inputs, variance-based SA expresses the output variance as a sum of contributions from isolated inputs and their interaction, or, more precisely, it is based on an analysis-of-variance (ANOVA) of the high-dimensional model representation (HDMR) decomposition~\cite{saltelli_global_2008, qian_multifidelity_2018}.

Consider a mathematical model $f$ with random inputs characterized by the vector $\bm{Z}:\Omega_{\bm{Z}}\to\mathbb{R}^{d}$ distributed according to the joint density $\rho_{\bm{Z}}(\bm{z})=\prod_{i=1}^{d}\rho(z_{i})$ and scalar output $Y\in\mathbb{R}$ with density $\rho(Y)$.
A typical interest in UQ studies is to characterize first and second statistical moments of $Y$, i.e.
\begin{equation}\label{eq:mean_variance_integral}
    \mu[Y] = \mathbb{E}[f(\bm{Z})] = \int_{\Omega_{\bm{Z}}}\,f(\mathbf{z})\, \rho_{\mathbf{Z}}(\mathbf{z})\,\text{d}\bm{z},\quad\text{and}\quad\mathbb{V}[Y] = \mathbb{V}[f(\mathbf{Z})] = \int_{\Omega_{\bm{Z}}} (f(\mathbf{z}) - \mu[Y])^2\,\rho_{\mathbf{Z}} (\mathbf{z})\,\text{d}\bm{z}.
\end{equation}
To approximate the integrals in~\eqref{eq:mean_variance_integral}, evaluations of the model on a set of $N$ samples, $\{ \mathbf{z}^{(s)}\}^{N}_{s=1}\sim\rho(\bm{Z})$, can be used to construct the unbiased estimators
\begin{equation}
    \hat{E} = \frac{1}{N} \sum_{s=1}^N f(\mathbf{z}^{(s)}) \quad \text{and} \quad \hat{V} = \frac{1}{N-1} \sum_{s=1}^N (f(\mathbf{z}^{(s)}) - \hat{E})^2.
    \label{eq:Mean_and_Variance}
\end{equation}

Additionally, the influence of $Z_{j}$ on the variance of output $Y$ can be assessed through the \emph{direct} Sobol' index $S_j$
\begin{equation}\label{eq:Main_Sobol}
    S_j = \frac{V_j}{V} = \frac{\mathbb{V} \, [\mathbb{E}[Y|Z_j]]}{\mathbb{V} [Y]}, \quad j= 1,...,d,
\end{equation}
which can be interpreted as the fraction of $\mathbb{V}[Y]$ \emph{explained} by the $j$-th input. The \emph{total} Sobol' index~$ST_j$ summarizes instead the \emph{combined} contributions to $\mathbb{V}[Y]$ from the $j$-th input plus the interactions with the other inputs, as 
\begin{equation}
    ST_j = \frac{T_j}{V} = \frac{\mathbb{V} [Y] - \mathbb{V} \, [\mathbb{E}[Y|\bm{Z}_{\sim j}]]}{\mathbb{V} [Y]} = 1 -\frac{\mathbb{V} \, [\mathbb{E}[Y|\bm{Z}_{\sim j}]]}{\mathbb{V} [Y]}, \quad j= 1,...,d,
    \label{eq:Total_Sobol}
\end{equation}
where $\bm{Z}_{\sim j}$ is the set of all inputs except $Z_{j}$. 
Interaction among the inputs is negligible in cases where $S_j \approx ST_j$.

Sobol' indices can be estimated efficiently with two independent sets of realizations from $\bm{Z}$, where the first set is denoted as $\{\mathbf{z}^{(A,s)} \}_{s=1}^N$ resulting in output $\{y^{(A,s)} \}_{s=1}^N$, and the second set is denoted $\{\mathbf{z}^{(B,s)} \}_{s=1}^N$ resulting in output $\{y^{(B,s)} \}_{s=1}^N$. 
These sets are then combined in such a way that the $j$-th component of the second set is replaced by the $j$-th component of the first set to form $\{\mathbf{z}^{(C_j,s)}\}_{s=1}^N$
where~\cite{saltelli_variance_2010}
\[
\mathbf{z}^{(C_j,s)} = 
\begin{bmatrix} 
z_1^{(B,s)} & \cdots & z_{j-1}^{(B,s)} &z_j^{(A,s)} & z_{j+1}^{(B,s)} & \cdots &z_N^{(B,s)}
\end{bmatrix}.
\]
Finally, the contribution of the $j$-th input to the output variance can then be approximated with a bias-corrected estimator~\cite{owen_variance_2013} as 
\begin{equation}
    \hat{V}_{j} = \frac{2N}{N-1} \left ( \frac{1}{N} \sum_{s=1}^{N} f \left (\mathbf{z}^{(A,s)} \right) f\left ({\mathbf{z}^{(C_j,s)}} \right) - \left ( \frac{\hat{E} + \hat{E}'}{2}\right)^2 + \frac{\hat{V} + \hat{V}'}{4N}\right ),
    \label{eq:V_Owen}
\end{equation}
where the realizations $\{y^{(A,s)} \}_{s=1}^N$ and $\{y^{(B,s)} \}_{s=1}^N$  are used to estimate $\hat{E}$, $\hat{E}'$ and $\hat{V}$, $\hat{V}'$, respectively. An unbiased estimator for $T_j$ is also given by~\cite{owen_variance_2013}
\begin{equation}
    \hat{T}_{j} = \frac{1}{2N} \sum_{i=1}^{N} \left ( f\left (\mathbf{z}^{(B,s)} \right) - f\left ( {\mathbf{z}^{(C_j,s)}} \right)\right ).
    \label{eq:T_Owen}
\end{equation}

\subsection{Multifidelity Monte Carlo estimation of Sobol' indices}

We follow the framework proposed in~\cite{qian_multifidelity_2018} for multi-fidelity Monte Carlo (MFMC) estimation of Sobol' sensitivity indices. In the multifidelity setting, Sobol' sensitivity indices are estimated using $K$ models, where the highest-fidelity model is denoted by $f^{(1)}$ and the models $f^{(k)}$ for $k = 2, ..., K$ decrease in fidelity with increasing~$k$.
An unbiased estimator of $\mathbb{V}[f^{(1)}(\mathbf{Z})]$ is the multifidelity variance estimator~$\hat{V}_{\text{mf}}$ expressed as
\begin{equation}
    \hat{V}_{mf} = \hat{V}_{m_1}^{(1)} + \sum_{k=2}^K \alpha_k \left ( \hat{V}_{m_k}^{(k)} - \hat{V}_{m_{k-1}}^{(k)}\right),
    \label{eq:MF_Variance_Estimator}
\end{equation}
where $\alpha_k,\,k=2\dots,K$ are control variate coefficients, and the subscripts ($\boldsymbol{m} = [m_1,..., m_{k-1}, m_k, ..., m_K] \in \mathbb{N}^K$) denote the set of samples (also referred to as \emph{allocation}) used for the evaluation of $f$. The variance of the multi-fidelity variance estimator in Equation~\eqref{eq:MF_Variance_Estimator} can be  expressed as \cite{qian_multifidelity_2018}
\begin{align}\label{eq:analytical_Variance}
    \begin{split}
        \mathbb{V}[\hat{V}_{mf}(\mathbf{Z})] = & \frac{1}{m_1}\left( \delta_1 - \frac{m_1 -3}{m_1-1} \sigma_1^4\right) + \sum_{k=2}^K \alpha_k^2 \left[ \frac{1}{m_{k-1}} \left( \delta_k - \frac{m_{k-1} - 3}{m_{k-1}-1} \sigma_k^4\right) - \frac{1}{m_k} \left( \delta_k - \frac{m_{k} - 3}{m_{k}-1} \sigma_k^4\right)\right] \\
    & + 2 \sum_{k=2}^{K} \alpha_k \left[ \frac{1}{m_k} \left( q_{1k} \tau_1 \tau_k + \frac{2}{m_k-1} \rho_{1k}^2 \sigma_1^2 \sigma_k^2\right) - \frac{1}{m_{k-1}} \left( q_{1k} \tau_1 \tau_k + \frac{2}{m_{k-1}-1} \rho_{1k}^2 \sigma_1^2 \sigma_k^2\right)\right],
    \end{split}
\end{align}
where $\delta_k$ is the fourth moment of $f^{(k)}(\mathbf{Z})$, $\tau_k$ is the standard deviation of $g^{(k)}(\mathbf{Z}) = (f^{(k)}(\mathbf{Z}) - \mathbb{E}[f^{(k)}(\mathbf{Z})])^2$, and $q_{k,l}~=~\mathbb{C}ov[g^{(k)}(\mathbf{Z}),g^{(l)}(\mathbf{Z})]/(\tau_k\,\tau_l)$.

With the two independent sets of realizations~$f(z^{A,s})$ and $f(z^{B,s})$, the multifidelity estimators for $\hat{V}_j$ and $\hat{T}_j$ follow as 
\begin{equation}\label{eq:MF_Total_Sobol_Estimator}
        \hat{V}_{j,mf} = \hat{V}_{j,m_1}^{(1)} + \sum_{k=2}^K \alpha_k \left ( \hat{V}_{j,m_k}^{(k)} - \hat{V}_{j,m_{k-1}}^{(k)}\right),\,\,\text{and}\,\,\hat{T}_{j,mf} = \hat{T}_{j,m_1}^{(1)} + \sum_{k=2}^K \alpha_k \left ( \hat{T}_{j,m_k}^{(k)} - \hat{T}_{j,m_{k-1}}^{(k)}\right),
\end{equation}
using the estimators of~\eqref{eq:V_Owen}-\eqref{eq:T_Owen}.
Following the definition of the main and total Sobol' indices in~\eqref{eq:Main_Sobol} and~\eqref{eq:Total_Sobol}, their multifidelity estimators can be written as
\begin{equation}
    \hat{S}_{j,mf} = \frac{\hat{V}_{j,mf}}{\hat{V}_{mf}} \quad \text{and} \quad \widehat{ST}_{j,mf} = \frac{\hat{V}_{j,mf}}{\hat{V}_{mf}}.
    \label{eq:MF_Sobol_Estimators}
\end{equation}

The control variate coefficients $\alpha_k,\,k=2,\dots,K$ and \emph{allocations} $\bm{m}$ are determined by minimizing the mean squared error (MSE) of the multifidelity mean estimator~$\hat{E}_{mf}$
\begin{equation}
    \hat {E}_{mf} = \hat E_{m_1}^{(1)} + \sum_{k=2}^K \alpha_k \left( \hat E_{m_k} ^{(k)} - \hat E_{m_k} ^{(k-1)}\right),
\end{equation}
for a given computational budget~$p \in \mathbb{R}_+$. 
For $K$ models $f^{(1)}, ..., f^{(K)}$ with correlations of $|\rho_{1,1}| > ... > |\rho_{1,K}|$ and computational costs~$\mathbf{w} = [w_1, ..., w_K]$ satisfying 
\begin{equation}
    \frac{w_{k-1}}{w_k} = \frac{\rho_{1,k-1}^2- \rho_{1,k}^2} {\rho_{1,k}^2- \rho_{1,k+1}^2},
    \label{eq:Qian_theorem_3.5}
\end{equation}
optimal coefficients $\alpha_k,\,k=2,\dots,K$ and allocations $\bm{m}$ are evaluated with the standard deviation~$\sigma_k = \sigma[f^{(k)}(\bm Z)]$ as
\begin{equation}\label{eq:optimal_allocation}
\alpha_k = \frac{\rho_{1,k} \sigma_1}{\sigma_k},\,\,m_k = m_1 \cdot r_k = \frac{p}{\mathbf{w}^T\,\mathbf{r}}\cdot r_k,\,\,\text{where}\,\,\mathbf{r} = [r_1, ..., r_K],\,\,\text{and}\,\,r_k = \sqrt{\frac{w_1 (\rho_{1,k}^2 - \rho_{1,k+1}^2)}{w_k (1 - \rho_{1,2}^2)}}.
\end{equation}
The statistics $\rho_{1,k}$ and $\sigma_k$ are determined from model results corresponding to a limited number of randomly selected input locations referred to as a \emph{pilot run}. 
It has been shown that this approach of model allocation is robust for estimation of variances and sensitivity indices~\cite{peherstorfer_optimal_2016}.
A more general approach is to solve an optimization problem that minimizes the MSE of the multifidelity variance estimate~$\mathbb{V}[\hat{V}_{mf}]$. 

Note that in order to estimate sensitivity indices with the two sets of realizations $\{y^{(A,s)} \}_{s=1}^N$ and $\{y^{(B,s)} \}_{s=1}^N$, a total of $d+2$ model evaluations are needed. Therefore, an effective computational budget $p_{\text{eff}} = p/(d+2)$ is distributed across all fidelity levels.

\subsection{Polynomial Chaos estimation of Sobol' indices}

The model output~$Y$ is represented in PC by a pseudo-spectral expansion in terms of polynomials that are chosen to be orthogonal with respect to the distributions of the inputs $\bm{Z}$~\cite{eck_guide_2016}. 
Model output and model output variance can then be approximated with $N+1$ orthogonal polynomials~$\Phi_p,\,p=0\dots,N$ as
\begin{equation}
    Y \approx Y_{PC} \approx \sum_{p=0}^{N} c_p \, \Phi_p (\mathbf{Z}), \quad \text{and} \quad \mathbb{V}[Y] \approx \mathbb{V}[Y_{PC}] = \sum_{p=0}^N \mathbb{V}[c_p \, \Phi_p(\textbf{Z})],
\end{equation}
where $c_p$ are expansion coefficients computed in this work through least-squares regression to minimize the $L^2$-norm difference between the PC expansion and a set of model evaluations sampled from the Sobol' sequence. 
These coefficients are determined from the solution of the overdetermined linear system resulting from using at least twice as many samples as the total number of coefficients~$c_p$.
Statistical moments and sensitivity measures can be derived analytically from the PC expansion coefficients following \cite{sudret_global_2008}. The main Sobol' index~$S_i$~\cite{saltelli_global_2008}, is computed from the PC expansion as the ratio of the output variance due to $z_i$ to the total model output variance:

\begin{equation}
S_i \approx \frac{1}{\mathbb{V}[Y_{PC}]} \sum_{p \in A_i} \mathbb{V}[c_p \hspace{1mm} \Phi_p],
\end{equation}

where the set $A_i$ contains all basis functions solely dependent on $z_i$. To evaluate the impact of interactions among model parameters, the variance of parameter $z_i$ and its interactions with $z_{\sim i}$ are related to the total model output variance through the total Sobol' index $ST_i$ as:

\begin{equation}
ST_i \approx \frac{1}{\mathbb{V}[Y_{PC}]} \sum_{p \in A_{T,i}} \mathbb{V}[c_p \hspace{1mm} \Phi_p],
\end{equation}
where $A_{T,i}$ represents the set of all basis functions depending on $z_i$.

\subsection{Numerical models of common carotid artery haemodynamics}

\subsubsection{Three-dimensional FSI model}

Blood flow within the three-dimensional vessel lumen domain~$\Omega_f$ is governed by the incompressible Navier-Stokes equations expressed as
\begin{equation}
\begin{split}
\nabla \cdot \bm{u} = 0, & \quad \bm{x} \in \Omega_f^{3D}, \quad t \in \mathbb{R}_{\geq 0}, \\
\rho_f \frac{\partial \bm{u}}{\partial t} + \rho_f (\bm{u} \cdot \nabla)\,\bm{u} = - \nabla P + \eta \, \Delta \bm{u} + \textbf{f}, & \quad \bm{x} \in \Omega_f^{3D}, \quad t \in \mathbb{R}_{\geq 0},
\end{split}
\end{equation}
with the velocity field $\bm{u}(\bm{x},t)$, the pressure field $P(\bm{x},t)$, fluid density $\rho_f$, the dynamic viscosity~$\eta$, and body forces~$\bm{f}(\bm{x},t)$. 
Initial conditions are prescribed as $\bm{u}(\bm{x}, t) = \bm{u}_0 (\bm{x})$ and $p(\textbf{x}, t) = p_0 (\textbf{x})$ for $\textbf{x} \in \Omega_f^{3D}$, and a time-dependent flow rate at the inlet~$\Gamma_{in}$ is specified through Dirichlet boundary conditions of the form
\begin{equation}
    \bm{u}(\bm{x},t) = \bm{u}_{in} (t), \quad x \in \Gamma_{in}, t \in \mathbb{R}_{\geq 0},
\end{equation}
while flow rate and outlet pressure at outlet $\Gamma_{out}$ are related by algebraic-differential equations
\begin{equation}
    P_{out}(\bm{x},t) = f(\bm{x}, t, Q_{out}(t), \dot Q_{out}(t), \bm{\phi}),\quad \bm{x} \in \Gamma_{out}, t \in \mathbb{R}_{\geq 0}. 
\end{equation}
The quantity $\bm{\phi}$ represents a set of parameters characterizing the haemodynamic response of the downstream vasculature. At the fluid-structure interface~$\Gamma_w$, a no-slip boundary condition is prescribed by
\begin{equation}
    \bm{u} (\bm{x,t})  = \bm{0}, \quad x \in \Gamma_w, t \in \mathbb{R}_{\geq 0}.
\end{equation}

The equilibrium in the structural domain~$\Omega_s$ is enforced through the Cauchy momentum equation as
\begin{equation}
    \begin{split}
        \rho \frac{\partial \bm{u}}{\partial t} &  = \nabla \cdot \sigma + \rho\,\bm{f}_b \\
        \sigma \cdot \bm{n}_s &= \bm{t}, \quad \text{on}\, (\Gamma_t)_h,
    \end{split}
\end{equation}
where $\sigma$ represents the Cauchy stress tensor, and $\textbf{n}_s$, $\textbf{t}$ denote the surface normal vector and boundary traction, respectively.

Vascular tissue is simulated using a neo-Hookean material model, where the second Piola-Kirchhoff stress~$\textbf{S}$ can be written as
\begin{equation}\label{equ:constitutive}
\bm{S} = \bm{S}_{0}+\bm{S}_{iso} + \bm{S}_{vol},\,\,\text{with}\,\,\bm{S}_{iso} = \mu J^{-2/3} \left ( \bm{I} - \frac{1}{3} \text{tr}(\bm{C}) \bm{C}^{-1}\right )\,\,\text{and}\,\,\bm{S}_{vol} = p J \bm{C}^{-1}.
\end{equation}
In~\eqref{equ:constitutive}, $\bm{S}_{0}$, $\bm{S}_{iso}$ and $\bm{S}_{vol}$ denote an initial prestress, the isocoric and volumetric component of $\bm{S}$, respectively. 
In addition, $\mu = E/[2(1+\nu)]$ is the second Lam\'e constant, $E$ is the elastic modulus, and $\nu$ is the Poisson ratio. The quantity $J$ represents the Jacobian of the deformation tensor $\bm{F} = \partial\bm{x}/\partial \bm{X}$, and $\textbf{I}$, $\textbf{C} = \textbf{F}^T \textbf{F}$ stand for the identity tensor and right Cauchy-Green deformation tensor, respectively. Spatial and material coordinates are denoted by $\partial\bm{x}$ and $\partial \bm{X}$.

The coupling between fluid and solid follows an Arbitrary Lagrangian-Eulerian formulation~\cite{bazilevs_isogeometric_2008}, where the meshes inside of each domain can deform, nodes on the fluid-solid interface coincide, and pressures, velocities, and displacements are uniquely defined at these locations. 

In our work, the common carotid artery (CCA) is modelled as an idealized straight cylinder. An unstructured mesh for each domain was generated in the open source 3D finite element mesh generator Gmesh \cite{geuzaine_gmsh_2009}. 
The svFSI open-source finite element solver \cite{zhu_svfsi_2022} from the SimVascular software platform~\cite{updegrove_simvascular_2017, lan_re-engineered_2018} was used to conduct the three-dimensional FSI simulations using a gernealized minimum residual (GMRES) solver with diagonal preconditioner from the Trilinos Project~\cite{heroux_overview_2005,seo_performance_2019}.
The structural solver was modified to account for an initial prestress~$\mathbf{S}_0$ superimposed to the second Piola-Kirchhoff stress~$\mathbf{S}$~\cite{hsu_blood_2011}. 
Average fluid tractions over one cardiac cycle were computed from a separate rigid-wall fluid dynamics simulation, and $\mathbf{S}_0$ was iteratively determined to produce a zero pre-stressed displacement response, i.e., equilibrating those tractions. Successive time-dependent FSI simulations used this equilibrium configuration as an initial condition.
Finally, time integration is performed with the generalized-$\alpha$ method~\cite{jansen2000generalized}, and all simulations ran on high performance computing resources available through the Center for Research Computing at the University of Notre Dame.

\subsubsection{One-dimensional model}

\begin{figure}[!ht]
	\centering
	\vspace{1.5mm}
	\begin{overpic}[abs,unit=1mm,scale=.28]{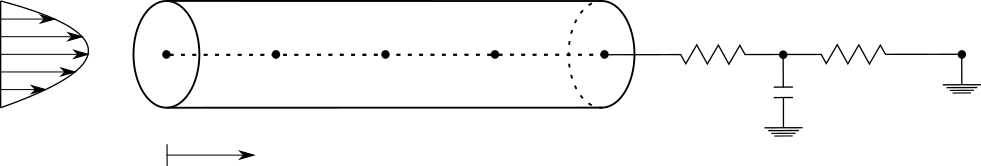}
		\put(62,10){\scriptsize$R_d$}
		\put(52,10){\scriptsize$R_p$}
		\put(55,5){\scriptsize$C$}
		\put(-9,7){\scriptsize$Q_{in}(t)$}
		\put(14.5,1.5){\scriptsize$z$}
	\end{overpic}
	\vspace{3mm}
	\caption{One-dimensional model for the CCA with prescribed parabolic inflow and three-element Windkessel outlet boundary conditions. The arterial impedance $Z$, compliance $C$, and resistance $R$ are used to characterize the response of the downstream vasculature.}
	\label{fig:StudyRepresentation}
\end{figure}
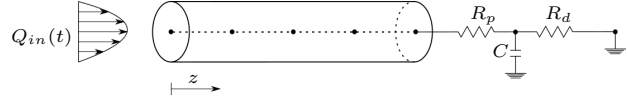

The CCA is modelled as a straight, deformable tube with a spatial discritization of five nodes and four elements along the center line. 
Blood is assumed to be an incompressible Newtonian fluid with density~$\rho_f$ and dynamic viscosity~$\eta$. Pressure~$P$ and flow rate~$Q$ are evaluated as cross-sectional averages at each node. The governing equations for the conservation of mass and momentum are
\begin{subequations}\label{eq:conservationLaws}
	
	\begin{align}
		\frac{\partial A}{\partial t} + \frac{\partial (A\,u)}{\partial z} & = 0, \quad z \in \Omega^{1D}, t \in \mathbb{R}_{\geq 0},\\
		\frac{\partial u}{\partial t} + u \frac{\partial u}{\partial z} + \frac{1}{\rho_f} \frac{\partial P}{\partial z} & = \frac{f}{\rho_f A}, \quad z \in \Omega^{1D}, t \in \mathbb{R}_{\geq 0},
	\end{align}
\end{subequations}
where $A$ is the cross-sectional area and $f$ represents a frictional force term per unit length. 
This frictional term accounts for wall shear stress and convective inertia. Its magnitude is determined by an axisymmetric polynomial velocity profile expressed as
\begin{equation}
	u_{r}(z, r, t) = u(z,t) \frac{\zeta + 2}{\zeta} \left[ 1- \left(\frac{r}{R}\right)^{\zeta} \right].
	\label{eq:velocityProfile}
\end{equation}
At a radial distance~$r$, the velocity~$u_r$ depends on the velocity profile shape determined through the polynomial order~$\zeta$ and the vessel radius~$R$. 
A laminar Poiseuille parabolic velocity profile $\zeta = 2$ leads to $f = -8\eta \pi u(z,t)$. 

The arterial wall is assumed to deform only in the circumferential direction and is modelled as a thin, homogenous, istotropic, elastic membrane that is impermeable and incompressible. 
Pressure and lumen cross-sectional area are then related by the ``tube law''~\cite{sherwin_one-dimensional_2003} as 
\begin{equation}\label{eq:constitutiveLaw}
P = P_{\mathrm{dia}} + \frac{\beta} {A_{\mathrm{dia}}} \left(\sqrt{A} - \sqrt{A_{\mathrm{dia}}}\right) \hspace{1cm} \mbox{with} \, \, \beta = \frac{\sqrt{\pi} E h}{\left(1-\nu^2\right) },	
\end{equation}
where $A_{\mathrm{dia}}$ and $P_{\mathrm{dia}}$ are the diastolic cross-sectional area and pressure, respectively. 
Arterial wall properties are the elastic modulus~$E$, wall thickness~$h$, and Poisson ratio~$\nu$.

The system~\eqref{eq:conservationLaws} is solved with a second order (in both space and time) explicit MacCormack scheme~\cite{boileau_benchmark_2015} and all simulations were performed with the open-source STARFiSh simulation program \cite{eck_uncertainty_2016} on a single processor of a desktop computer with Intel(R) Xeon(R) CPU E5-2630 V4 at 2.20GHz with 32GB RAM.
 
\subsubsection{Zero-dimensional model}
\begin{figure}[ht!]
    \centering
    	\begin{overpic}[abs,unit=1mm,scale=.45]{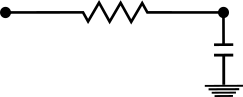}
		\put(13,6){\scriptsize$R$}
		\put(22,4.5){\scriptsize$C$}
	\end{overpic}
    \caption{Zero-dimensional RC model of the CCA.}
    \label{fig:0D_model}
\end{figure}
Lumped parameter or zero-dimensional haemodynamic models represent blood vessels as a network of conduits where the spatial averages of pressure~$P$ and volumetric flow~$Q$ evolve with time.
These models consist of electrical circuit elements, where the hydrodynamic analog of electrical current and voltage are flow rate and pressure, respectively. 
Viscous effects are modeled through resistors, whereas capacitors and inductors mimic vessel deformability and blood inertia, respectively. These system are governed by algebraic-differential equations of the form
\begin{equation}
\Delta P = RQ, \quad Q = C\Delta\dot P.
\end{equation}
Vessel resistance and compliance can be evaluated under the assumption of an ideal cylindrical vessel geometry, laminar Poiseuille flow and linear elastic material properties of the vascular tissue as 
\begin{equation}
R = \frac{8 \eta L}{\pi r^4}, \quad C = \frac{3 L \pi r^3}{2Eh}, 
\end{equation}
where $\eta$ is the dynamic viscosity of blood, $L$ is the vessel length, $r$ is the diastolic vessel radius, $E$ is the elastic modulus, and $h$ is the vessel wall thickness.
In this work, the CCA was represented with two identical electrical circuit units comprising a resistance and compliance element as shown in Figure~\ref{fig:0D_model}. This set-up allowed determining QoIs at the longitudinal centroid of the vessel. 
Simulations were performed using the svZeroDPlus solver from the open-source SimVascular software platform. svZeroDPlus performs time integration with an implicit generalized-$\alpha$ method~\cite{jansen2000generalized}, and simulations were performed on a single processor of a desktop computer with Intel(R) Xeon(R) CPU E5-2630 V4 at 2.20GHz with 32GB RAM.

\subsection{Boundary conditions, uncertainties, and quantities of interest}\label{sec:bc_inputs_qoi}

A physiological flow rate waveform with a parabolic velocity profile from prior studies \cite{figueroa_coupled_2006} was prescribed at the inlet of all models. 
At the outlet, a three-element Windkessel model was used for the downstream vasculature, with pressure and flow related to the electrical circuit constants as
\begin{equation}
    \frac{\partial P}{\partial t} + \frac{P}{R_d C} = Q \left ( \frac{1}{C} + \frac{R_p}{C R_d}\right ) + R_p \frac{\partial Q}{\partial t}.
\end{equation}
For all model fidelities, the boundary condition parameters given in Table~\ref{tab:Deterministic_Parameters} were adopted.

\begin{table}[!ht]
    \scriptsize
    \centering
    \caption{Deterministic parameters of all model fidelities.}
    \begin{tabular}{l|c|c}
        \toprule
        Parameter & Value & unit \\
        \midrule
         vessel length $L$ & 0.126 & m \\
         blood density $\rho$ & 1050 & kg m$^{-3}$\\
         blood viscosity $\eta$ & 0.001 & mPa s \\
         Poisons' ratio $\nu$ & 0.49 & - \\
         peripheral resistance $R_p$ \cite{xiao_systematic_2014} & 2.4875 $\cdot 10^{8}$ & Pa s m$^{-3}$\\
         vessel compliance $C$ \cite{xiao_systematic_2014} & 1.3546 $\cdot 10^{-10}$ & Pa s m$^{-3}$\\
         distal resistance $R_d$ \cite{xiao_systematic_2014} & 1.8697 $\cdot 10^{9}$ & Pa s m$^{-3}$\\
         \bottomrule
    \end{tabular}
    \label{tab:Deterministic_Parameters}
\end{table}

\begin{table}[!ht]
    \scriptsize
    \centering
    \caption{Mean values and ranges~[$a_i, b_i$] of the uncertain model parameters for SA for all model fidelities.}
    \begin{tabular}{l|c|c|c}
        \toprule
        Parameter & mean & range & unit \\
        \midrule
         vessel radius $r$ & 3.289 & [2.96, 3.62] & mm \\
         elastic modulus $E$ & 440 & [396, 484] & kPa\\
         wall thickness $h$ & 0.785 & [0.7065, 0.8635] & mm \\
         \bottomrule
    \end{tabular}
    \label{tab:Uncertain_Parameters}
\end{table}

Uncertainty was considered in the vessel radius~$r$, wall thickness~$h$, and elastic modulus~$E$. These uncertainties were based on values representing the CCA of a human in their mid forties, with a uniform distribution varying 10\% around the mean value~$\mu_i$; thus lower bound~$a_i$ and upper bound~$b_i$ were $0.9 \mu_i$ and $1.1 \mu_i$, respectively, as shown in Table~\ref{tab:Uncertain_Parameters}. 
Additionally, the QoIs selected for SA were the systolic pressure~$P_{\text{sys}}$, the pulse pressure~PP and the maximal radius change $\Delta r_{\text{max}}$, evaluated as a cross-sectional average at the 3D-FSI and 1D model mid-span, and the 0D model half-length.

\begin{figure}[!ht]
    \centering
    \includegraphics[width=1.0\textwidth]{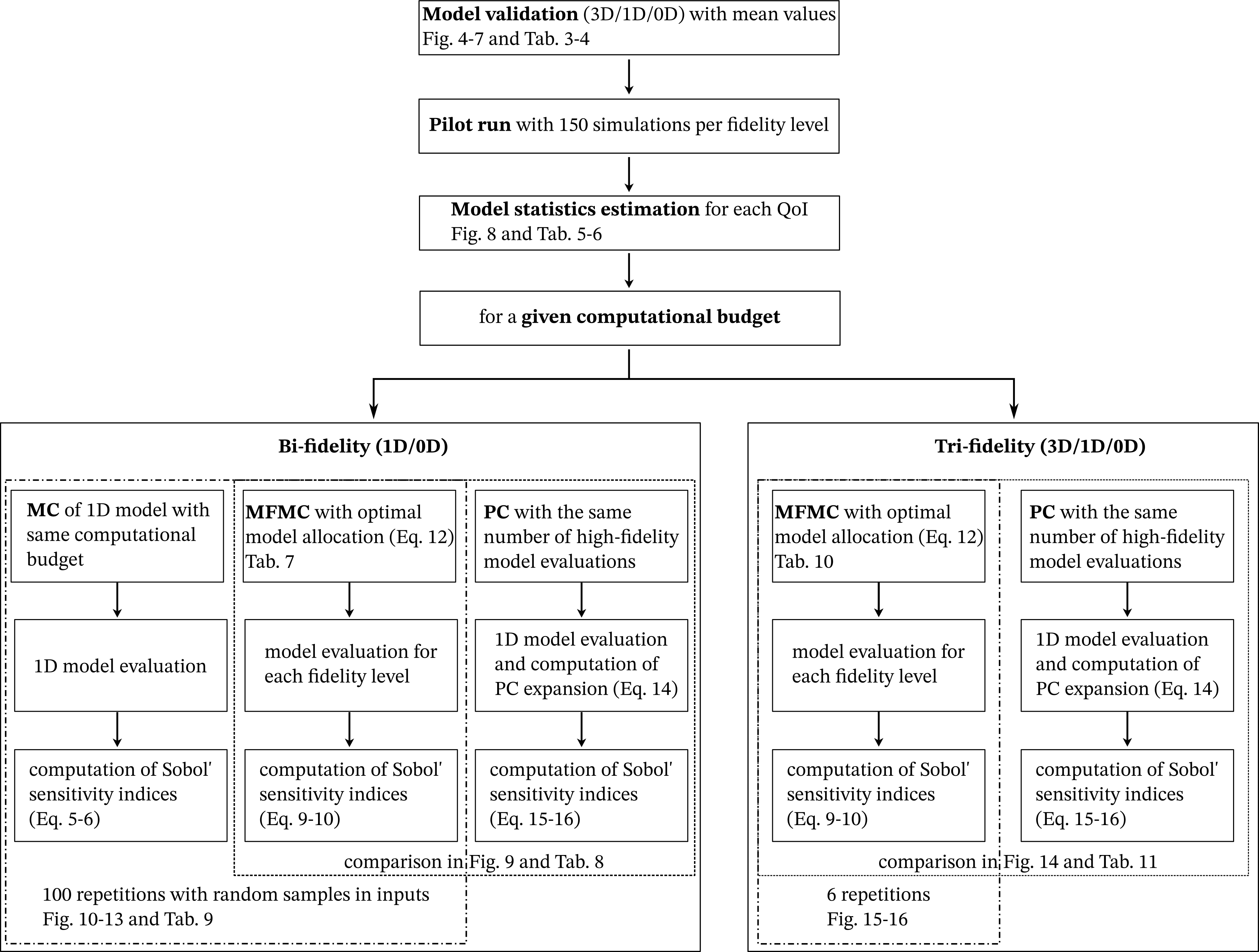}
    \caption{Computational pipeline for bi- and tri-fidelity MFMC estimators and comparative analysis with MC and PC results.}
    \label{fig:simulationPipeline}
\end{figure}

\section{Results}\label{sec:results}

In this section, we present the results of MFMC SA, using estimators for sentivitiy indices based on two and three model fidelities.
We first investigated how the selected QoIs were affected by the spatial and temporal discretization of the various models, and the number of cardiac cycles needed to obtain a periodic solution. For brevity, only results for the 3D-FSI model are reported. All subsequent simulations included in the study were set up based on the results of this preliminary validation.
A spatial discretization of 5~nodes and a time step of 0.0025\,s were selected for the 1D-model, resulting in a maximum Courant number of 0.8. The time step in the 0D-model was instead set to 0.001\,s. A total of five and ten cardiac cycles were evaluated for the 1D and 0D model, respectively.
A pilot run of 150~simulations for each fidelity was used to estimate model statistics and correlations between the models. 
Optimal model allocations were then evaluated from these pilot runs and used to compute multifidelity estimates of sensitivity metrics. 
Figure~\ref{fig:simulationPipeline} offers a comparative summary for the proposed simulation pipeline which includes MFMC, MC and PC estimators.

\subsection{Model validation}

\begin{figure}[!ht]
    \centering
    \includegraphics[width=1.0\textwidth]{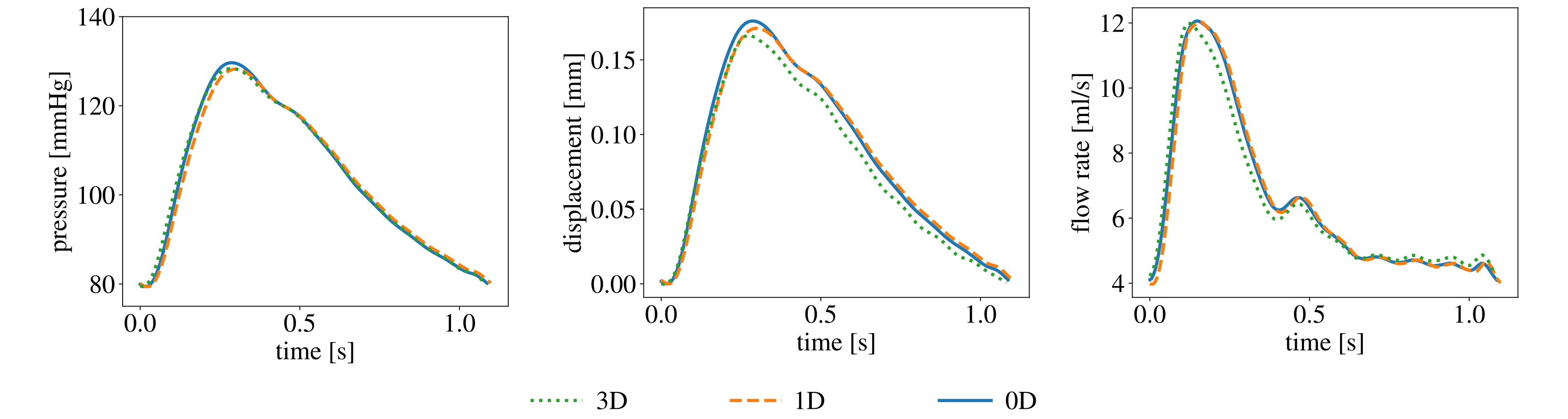}
    \caption{Illustration of mid-span pressure ($P$), radial displacement at the lumen ($\Delta r$), and flow rate ($Q$) evolution over one cardiac cycle, comparing 0D, 1D, and 3D-FSI models.}
    \label{fig:FidelityValidation}
\end{figure}

The three model fidelities were simulated at the mean value of the random inputs to ensure sufficient agreement in the predicted pressures~$P$, displacements~$\Delta r$ and flow rates~$Q$.
Figure~\ref{fig:FidelityValidation} displays these QoIs for one cardiac cycle and each model fidelity. 
Minor differences in the vessel displacement and flow rate can be observed between the 3D-FSI and the two lower fidelity models. Displacement amplitude and flow rate appear slightly lower in the 3D-FSI model.
%
Quantitative error measures are provided in Table~\ref{tab:Error_FidelityValidation} and were computed as~\cite{pfaller_automated_2022, xiao_systematic_2014}
\begin{align}
    \epsilon_P^{\text{avg}} & =  \frac{\sum_{i=1}^{N} \left |P_i^{\mathrm{3D}} - P_i^{\text{dD}} \right |}{\sum_{i=1}^{N}P_i^{\text{3D}}}, & 
    \epsilon_Q^{\text{avg}} & = \frac{1}{N} \frac{\sum_{i=1}^{N} \left |Q_i^{\mathrm{3D}} - Q_i^{\text{dD}} \right |}{\text{max}(Q_i^{\text{3D}}) - \text{min}(Q_i^{\text{3D}})}, & 
    \epsilon_{\Delta r}^{\mathrm{avg}} & = \frac{1}{N} \frac{\sum_{i=1}^{N} \left |\Delta r_i^{\mathrm{3D}} - \Delta r_i^{\text{dD}} \right |}{\text{max}(\Delta r_i^{\text{3D}}) - \text{min}(\Delta r_i^{\text{3D}})}, \nonumber \\
    \epsilon_P^{\text{max}}  &= \frac{\max_{i} \left |{P}_i^{\text{3D}} - {P}_i^{\text{dD}}\right|} {\frac{1}{N} \sum_{i=1}^{N}{P}_i^{\text{3D}}}, & 
    \epsilon_Q^{\text{max}} &= \frac{\max_{i} \left |{Q}_i^{\text{3D}} - {Q}_i^{\text{dD}}\right|} {\frac{1}{N} \sum_{i=1}^{N}{Q}_i^{\text{3D}}}, & 
    \epsilon_{\Delta r}^{\mathrm{max}} &= \frac{\max_{i} \left |{\Delta r}_i^{\text{3D}} - {\Delta r}_i^{\text{dD}}\right|} {\frac{1}{N} \sum_{i=1}^{N}{\Delta r}_i^{\text{3D}}}, \nonumber \\
    \epsilon_P^{\mathrm{sys}}& = \frac{\left | \max (P^{\text{3D}}) - \max (P^{\text{dD}}) \right |}{\max (P^{\text{3D}})}, & 
    \epsilon_Q^{\mathrm{sys}}& = \frac{\left |\max (Q^{\text{3D}}) - \max (Q^{\text{dD}}) \right |}{\max (Q^{\text{3D}})},   &
    \epsilon_{\Delta r}^{\mathrm{sys}}& = \frac{\left | \max (\Delta r^{\text{3D}}) - \max (\Delta r^{\text{dD}}) \right |}{\max (\Delta r^{\text{3D}})},  \nonumber\\
    \epsilon_P^{\mathrm{dia}} &= \frac{\left |\min (P^{\text{3D}}) - \min (P^{\text{dD}}) \right |}{\frac{1}{N} \sum_{i=1}^{N}{P}_i^{\text{3D}}}, & 
    \epsilon_Q^{\mathrm{dia}} &= \frac{\left | \min (Q^{\text{3D}}) - \min (Q^{\text{dD}}) \right |}{\text{max}(Q_i^{\text{3D}}) - \text{min}(Q_i^{\text{3D}})}, &
    \epsilon_{\Delta r}^{\mathrm{dia}} &= \frac{\left | \min (\Delta r^{\text{3D}}) - \min (\Delta r^{\text{dD}}) \right |}{\text{max}(\Delta r_i^{\text{3D}}) - \text{min}(\Delta r_i^{\text{3D}})}. 
    \label{eq:error_measures}
\end{align}

Relative errors between the 3D-FSI and the 1D model were overall bigger than the errors between the 3D/0D model and the 1D/0D model. 
Smallest average errors were obtained for the pressure (0.52\,\%-1.15\,\%) while the average error for radial displacement and flow rate was significantly higher (1.63\,\%-5.08\,\%). 
Diastolic quantities showed overall smaller errors than systolic results.

\begin{table}[!ht]
    \scriptsize
    \centering
    \caption{Relative error measures from~\eqref{eq:error_measures} for each fidelity and QoI.}
    \begin{tabular}{c|c c c|c c c|c c c}
        \toprule
        & \multicolumn{3}{c}{pressure} & \multicolumn{3}{c}{displacement} & \multicolumn{3}{c}{flow rate} \\
        & $\epsilon_P^{3D/1D}$ & $\epsilon_P^{3D/0D}$ & $\epsilon_P^{1D/0D}$ & $\epsilon_{\Delta R}^{3D/1D}$ & $\epsilon_{\Delta R}^{3D/0D}$ & $\epsilon_{\Delta R}^{1D/0D}$ & $\epsilon_Q^{3D/1D}$ & $\epsilon_Q^{3D/0D}$ & $\epsilon_Q^{1D/0D}$ \\
        \midrule
        $\epsilon_{\text{QoI, avg}}$ [\%] & 1.15 & 0.52 & 1.00 & 5.04 & 4.42 & 2.14 & 5.08 & 3.78 & 1.63 \\
        $\epsilon_{\text{QoI, max}}$ [\%] & 5.29 & 2.98 & 0.86 & 13.4 & 3.28 & 3.92 & 27.1 & 13.8 & 3.98 \\
        $\epsilon_{\text{QoI, sys}}$ [\%] & 0.13 & 1.00 & 1.14 & 2.8 & 5.75 & 2.86 & 0.28 & 0.51 & 0.23\\
        $\epsilon_{\text{QoI, dia}}$ [\%] & 0.05 & 0.00 & 0.05 & 0.15 & 0.15 & 0.00 & 0.15 & 1.57 & 1.60\\
         \bottomrule
    \end{tabular}
    \label{tab:Error_FidelityValidation}
\end{table}

\begin{figure}[!ht]
    \centering
    \includegraphics[width=1.0\textwidth]{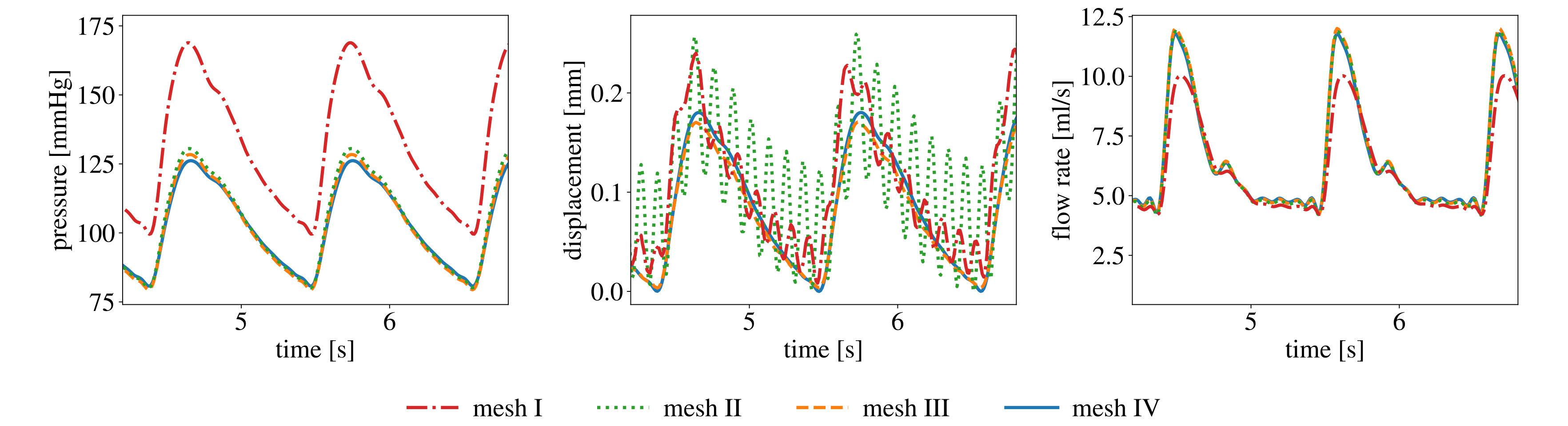}
    \caption{Mesh independence study results at artery mid-span.}
    \label{fig:3D_MeshIndependence}
\end{figure}

\begin{figure}[!ht]
    \centering
    \includegraphics[width=1.0\textwidth]{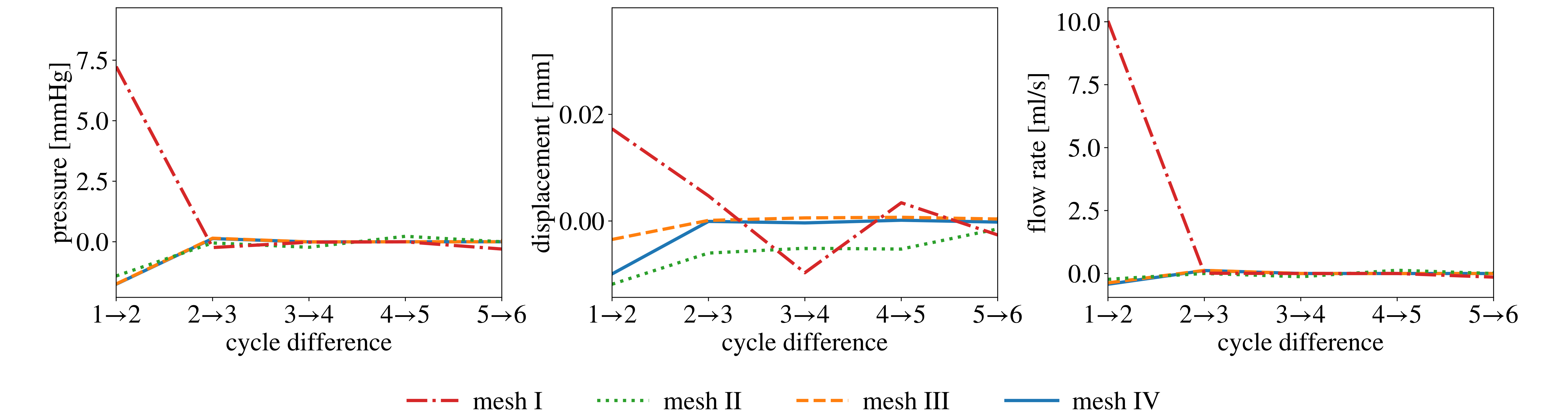}
    \caption{Investigating the number of simulated heart cycle leading to a periodic response for cross-sectional averages of pressure, displacement, and flow rate at the artery mid-span.}
    \label{fig:3D_CycleIndependence}
\end{figure}

\begin{figure}[!ht]
    \centering
    \includegraphics[width=1.0\textwidth]{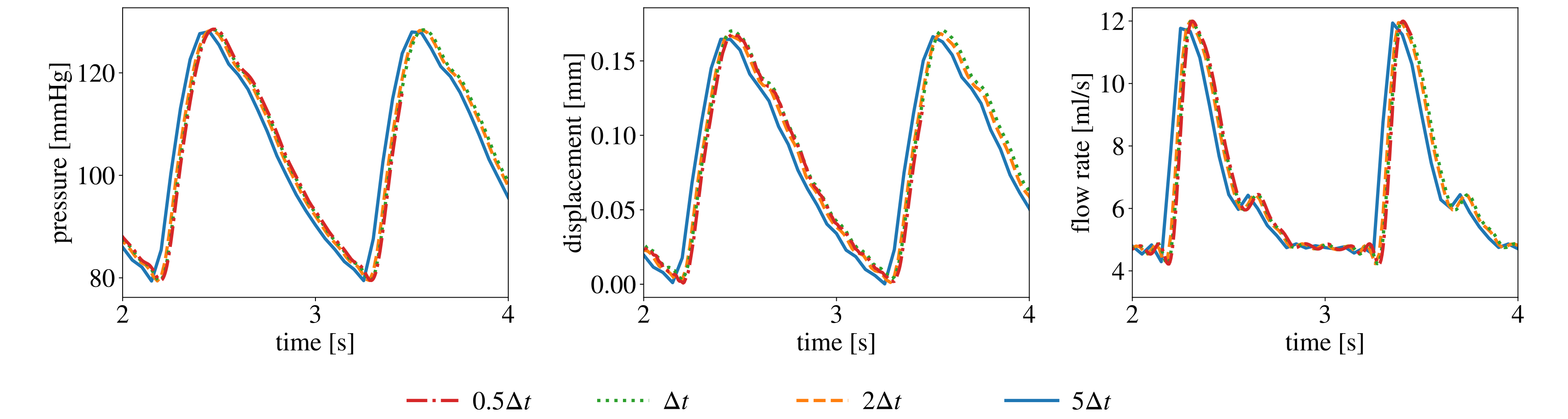}
    \caption{Time step independence study for mesh III.}
    \label{fig:3D_TimeIndependence}
\end{figure}

For the 3D-FSI model, mesh, time, and cycle independence studies were also conducted, with results presented in Figures~\ref{fig:3D_MeshIndependence}-\ref{fig:3D_TimeIndependence}. 
Four meshes were investigated ranging from 8,329 to 500,824 elements. 
Table~\ref{tab:MeshElements} shows detailed mesh statistics for each refinement, grouped by domain. For the wall mesh, the number of elements in the circumferential direction and the wall thickness are also given. 
As shown in Figure~\ref{fig:3D_MeshIndependence}, pressure and flow rate are approximately the same from the second mesh refinement. 
Radial displacement shows oscillations for mesh I and II, and are approximately the same and oscillation free for mesh III and IV. The average error between mesh III and IV for pressure, displacement, and flow rate were 1.14\,\%, 3.82\,\%, and 1.04\,\%, respectively. 
The best trade off between accuracy and computational cost was found for mesh III, therefore this mesh was adopted for all subsequent runs.

\begin{table}[!h]
    \scriptsize
    \centering
    \caption{Statistics for the total number of elements in mesh independence study.}
    \begin{tabular}{l|l|c|c|c|c}
        \toprule
         \multirow{2}{*}{\textbf{Mesh}} & \textbf{Total} & \textbf{Lumen} & \multicolumn{3}{c}{\textbf{Wall}}\\
          & & total & total & circumference & thickness \\
         \midrule
         mesh I & 8,329 & 2,908 & 5,421 & 28 & 1\\
         mesh II & 42,510 & 21,536 & 20,974 & 68 & 1\\
         mesh III & 241,466 & 157,839 & 83,627 & 184 & 2\\
         mesh IV & 500,824 & 295,976 & 204,857 & 300 & 2\\
         \bottomrule
    \end{tabular}
    \label{tab:MeshElements}
\end{table}

Pressure and flow converged to a periodic solution after the second cycle for all meshes, as shown in Figure~\ref{fig:3D_CycleIndependence}. Conversely, displacement converged after the second cycle for mesh III and IV, but only after the 6$^{\text{th}}$ cycle for mesh I and II. 
In addition, the difference between cycle three and four for mesh III was 0.16\,\%, 0.42\,\%, and 0.39\,\% for pressure, displacement, and flow rate, respectively. Therefore, a total of three cardiac cycles were simulated in all further 3D-FSI model evaluations.

Four different time steps were tested to investigate temporal convergence. The initial time step $\Delta t~=~0.0001$\,s was halved ($0.5 \Delta t$), doubled~($2\Delta t$), and multiplied by five~($5\Delta t$).
As shown in Figure~\ref{fig:3D_TimeIndependence}, a coarse time step of $5\Delta t$ shifted the pressure, displacement, and flow rate waveform to the left compared with the other time step sizes, which lead to the approximately same waveform and timing. 
The average difference in pressure, displacement, and flow rate between $\Delta t$ and $2 \Delta t$ was 1.6\,\%, 2.1\,\% and 2.2\,\% and  between $\Delta t$ and $0.5 \Delta t$ was 0.5\,\%, 1.4\,\% and 0.7\,\%, respectively. 
In all following 3D-FSI simulations, the time step was set to $\Delta t = 0.0001$\,s.

\subsection{Estimated model statistics from pilot runs}
From a pilot run with 150 simulations, the mean~$\mu$, standard deviation~$\sigma$, and correlation coefficient~$\rho_{3D,dD}$, $\rho_{1D,dD}$ were estimated for the three selected QoIs. 
The resulting 1D/0D and 3D/1D/0D model statistics are presented in Table~\ref{tab:ModelStatistics_1D_0D} and Table~\ref{tab:ModelStatistics_3D_1D_0D}, respectively. In the tri-fidelity case, model statistics are given only for P$_{\mathrm{sys}}$, while the values for all QoIs is given in the Appendix in Table~\ref{tab:ModelStatistics_3D_1D_0D_extended}.
The relative costs~$w$ of a single model evaluation are also reported for each model fidelity, quantified in terms of an equivalent number of high fidelity simulations.

\begin{table}[!ht]
    \scriptsize
    \centering
    \caption{Model response statistics for bi-fidelity estimator computed from 150 pilot runs. The statistics are the mean~$\mu$, the standard deviation~$\sigma$, the correlation with the 1D model~$\rho_{1D,dD}$, higher order statistics $\delta_k$, $\tau$, q, and the computational costs $w$.} 
    \begin{tabular}{l|c c c c|c c c c|c c c c}
        \toprule
         & \multicolumn{4}{c}{P$_{\mathrm{sys}}$} &\multicolumn{4}{c}{PP}  & \multicolumn{4}{c}{$\Delta r_{\text{max}}$} \\ 
        & 1D & 0D & 0D pert. & unit & 1D & 0D & 0D pert. & unit & 1D & 0D & 0D pert. & unit \\
        \midrule
            $\mu$ &  129.7 & 130.7 & 130.8 & mmHg & 51.1 & 51.9 & 51.9 & mmHg & 0.162 & 0.163 & 0.163 & mm\\
            $\sigma$ &  1.80 & 1.84 & 1.88 & mmHg & 2.80  & 2.80 & 2.81 & mmHg & 0.014 & 0.014 & 0.015 & mm \\
            $\rho_{3D,dD}$ &   1 & 0.9996 & 0.9986 & - & 1 & 0.9998 & 0.9994 & - & 1 & 0.9999 & 0.9999 & - \\
            $\delta_k$ &  23.42 & 24.77 & 26.82 & mmHg$^4$ & 138.6 & 137.2 & 137.2 & mmHg$^4$ & 1.00$\cdot 10^{-7}$  & 1.08$\cdot 10^{-7}$ & 1.15$\cdot 10^{-7}$  & mm$^4$\\
            $\tau_k$ &  474.0 & 484.8 & 500.1 & mmHg & 1159.1 & 1147.5 & 1141.9 & mmHg & 2.44$\cdot 10^{-7}$ & 2.53$\cdot 10^{-7}$ & 2.62$\cdot 10^{-7}$ & mm\\
            $q_{3D,dD}$ &  1 & 0.9991 & 0.9965  & - & 1 & 0.9996 & 0.9986 & - & 1 & 0.9998 & 0.9997 & -\\
         \midrule
         $w$ & 1 & 0.3 & 0.3 & - & 1 & 0.3 & 0.3 & - & 1 & 0.3 & 0.3 & - \\
         \bottomrule
    \end{tabular}
    \label{tab:ModelStatistics_1D_0D}
\end{table}

\begin{table}[!ht]
    \centering
        \caption{Model response statistics of P$_{\mathrm{sys}}$ for tri-fidelity estimator from 10, 20, or 150 pilot runs. The statistics are the mean~$\mu$, the standard deviation~$\sigma$, the correlation with the 1D model~$\rho_{1D,dD}$, higher order statistics $\delta_k$, $\tau$, q, and the computational costs $w$.} 
        \scriptsize
    \begin{tabular}{l c |c c c c c}
        \toprule
        \# &\textbf{}  & \multicolumn{5}{c}{P$_{sys}$}  \\ 
        & & 3D & 1D & 0D & 0D pert. & unit \\
        \midrule

        \multirow{6}{*}{10} & $\mu$ &  129.4 & 129.6 & 130.7 & 130.7 & mmHg \\
        & $\sigma$ &  1.68 & 1.56 & 1.59 & 1.45 & mmHg \\
        & $\rho_{3D,dD}$ &  1 & 0.9826 & 0.9827 & 0.9650 & - \\
        & $\delta_k$ &  30.38 & 25.23 & 26.50 & 20.06 & mmHg$^4$ \\
        & $\tau_k$ &  527.9 & 487.7 & 498.3 & 498.3 & mmHg \\
        & $q_{3D,dD}$ &  1 & 0.9938 & 0.9935 & 0.9837 & - \\
         \midrule
         
        \multirow{6}{*}{20} & $\mu$ &  129.6 & 129.7 & 130.8 & 130.8 & mmHg \\
        & $\sigma$ &  1.98 & 1.84 & 1.87 & 1.78 & mmHg \\
        & $\rho_{3D,dD}$ &  1 & 0.9915 & 0.9919 & 0.9878 & - \\
        & $\delta_k$ &  32.86 & 27.46 & 28.91 & 23.02 & mmHg$^4$  \\
        & $\tau_k$ &  524.0 & 498.8 & 509.7 & 509.7 & mmHg  \\
        & $q_{3D,dD}$ &  1 & 0.986 & 0.9874 & 0.986 & - \\
         
         \midrule         
          \multirow{6}{*}{150} & $\mu$ &  129.5 & 129.7 & 130.8 & 130.8 & mmHg \\
          & $\sigma$ &  1.93 & 1.80 & 1.84 & 1.79 & mmHg \\
          & $\rho_{3D,dD}$ & 1 & 0.9813 & 0.9835 & 0.969 & - \\
          & $\delta_k$ &  28.96 & 23.42 & 24.84 & 23.70 & mmHg$^4$ \\
        & $\tau_k$ &  512.1 & 474.0 & 484.8 &  483.2 & mmHg \\
        & $q_{3D,dD}$ & 1.0 & 0.9408 & 0.9459 & 0.9132 & - \\
         \midrule
         & $w$ & 1 & $9\cdot 10^{-5}$ & $3\cdot 10^{-5}$ & $3\cdot 10^{-5}$ & - \\
         \bottomrule
    \end{tabular}
    \label{tab:ModelStatistics_3D_1D_0D}
\end{table} 

\begin{figure}[!ht]
    \centering
    \includegraphics[width=1.0\textwidth]{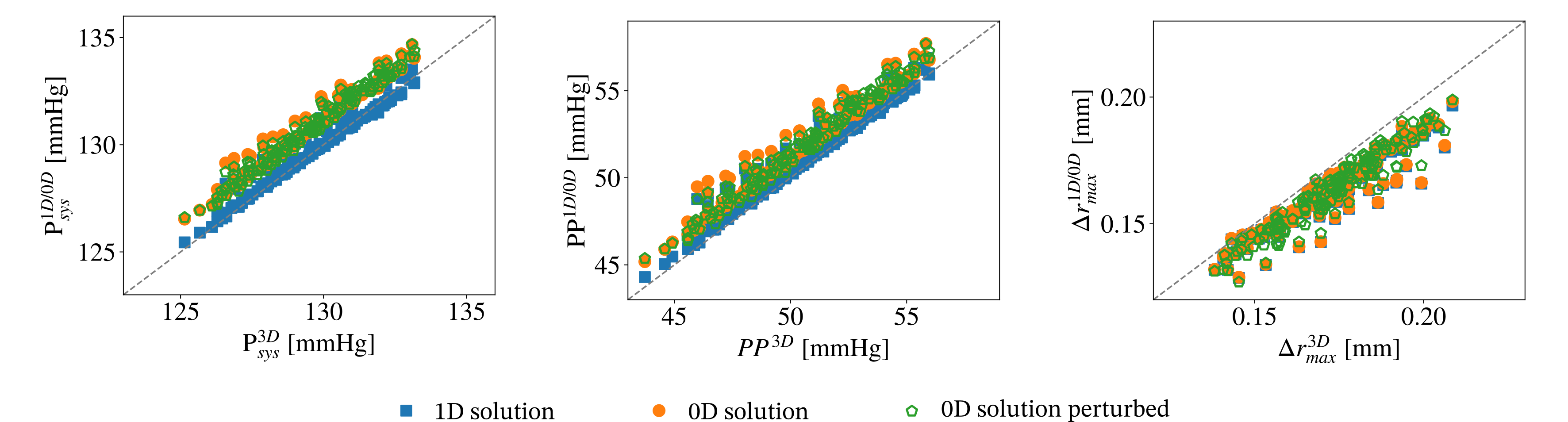}
    \caption{Scatter plot of low- vs. high-fidelity QoIs computed using 150 pilot runs.}
    \label{fig:EstimationStatistics}
\end{figure}

In this tri-fidelity set up, the correlation coefficient $\rho_{3D,1D}$ for P$_{\text{sys}}$ and PP was slightly smaller than $\rho_{3D,0D}$, or, in other words, the results from the 0D solver were found to be more correlated with 3D-FSI simulations outputs than 1D model results, even if 0D solutions were much cheaper.
For situations characterized by very high correlations of both low-fidelity models due to the idealized flow conditions considered in this study, this is not unexpected. 
Nevertheless, $\rho_{3D/1D} < \rho_{3D/0D}$ results in a deviation from the assumptions leading to Equation~\eqref{eq:Qian_theorem_3.5}, rendering it unsuitable for determining the optimal model allocations~$\textbf{m}$.
To reduce the correlation coefficient, the 0D model solution was perturbed through evaluation of the least squares discrepancy~$\textbf{d}$ between the 3D-FSI and 0D model parametric response. 
The perturbed 0D model was computed as
\begin{equation}\label{equ:0d_perturb}
Y^{0D}_{\mathrm{perturbed}}(\bm{z}^{(i)}) = Y^{0D}(\bm{z}^{(i)}) - \phi \, \cdot \textbf{d} s,
\end{equation}
where $\phi = 1$ is a perturbation factor, and $s$ is the sample matrix used to evaluate $Y^{0D}$. Samples of $s$ were drawn from the joint probability density function of the uncertain model parameters following the Sobol' sequence.
To see how the estimated model statistics vary with the number of pilot runs, tri-fidelity estimates were computed also for pilot runs consisting of 10 and 20 simulations (see Table~\ref{tab:ModelStatistics_3D_1D_0D} and \ref{tab:ModelStatistics_3D_1D_0D_extended}). 
Differences were less then 1\% for the mean values, up to 21.4\% or 7.1\% for the standard deviation and 5.6\% or 8.3\% for the correlation coefficient, for 10 and 20 pilot runs respectively. Higher statistical moments showed a maximal difference of 18.1-46.0\% for 10 pilot runs, while the maximal difference was between 7.1-24.0\% for 20 pilot runs. P$_\mathrm{sys}$ showed the least variations for a reduced number of pilot runs, while $\Delta r_{\mathrm{max}}$ was the most affected.

The low fidelity model results (i.e., 1D, 0D, and perturbed 0D model) from the pilot run are plotted versus the high fidelity model outputs in Figure~\ref{fig:EstimationStatistics}. 
This figure shows both an excellent agreement between the low- and high-fidelity model response, and the effect of the perturbation~\eqref{equ:0d_perturb} added to the 0D model outputs.
In the next sections, we compute bi-fidelity estimators of sensitivity indices using both unperturbed and perturbed 0D model results, while we only use perturbed 0D model results for tri-fidelity estimators.

\subsection{Results of 1D/0D bi-fidelity SA}

\begin{figure}[!ht]
    \centering
    \includegraphics[width=\textwidth]{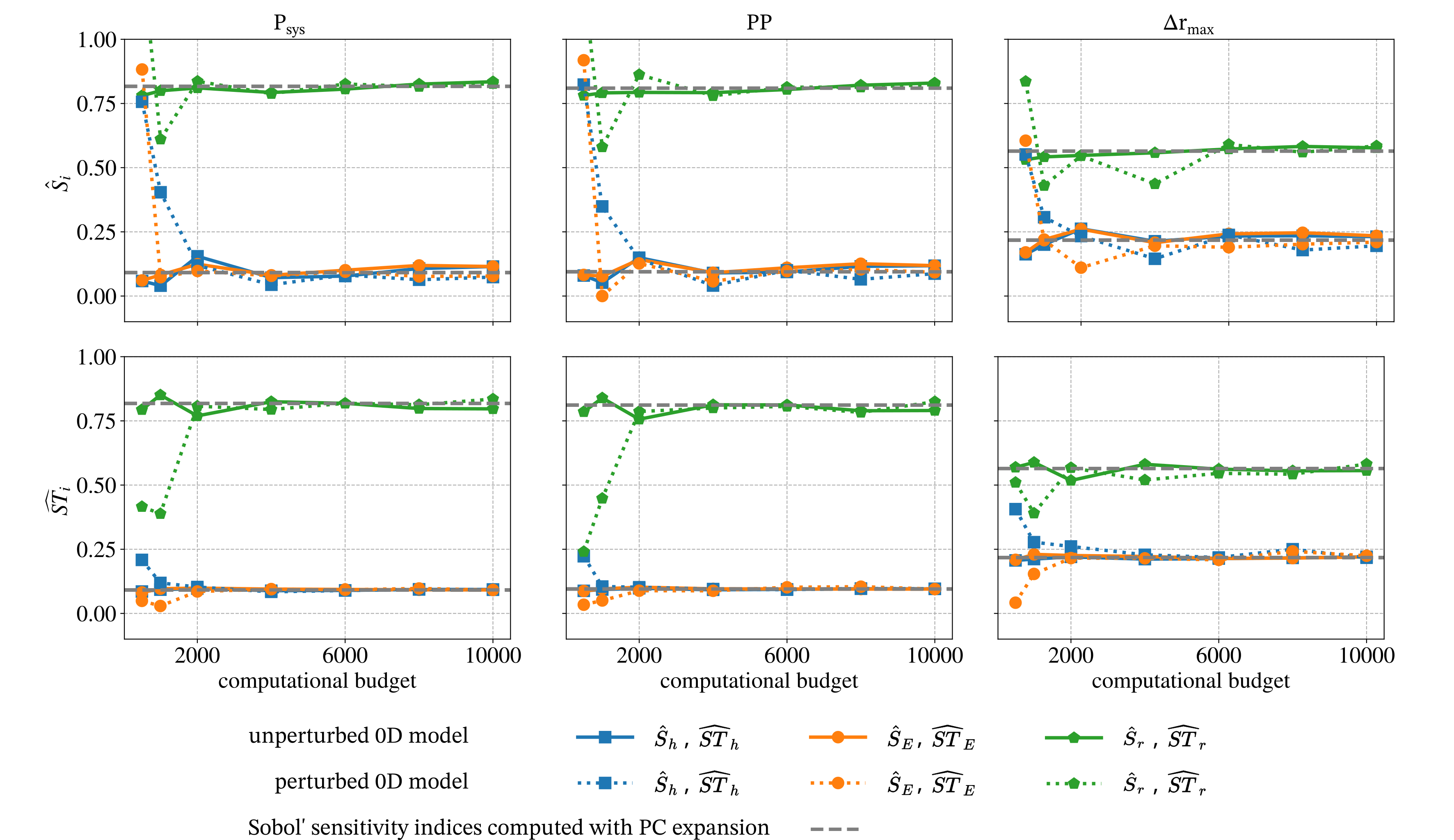}
    \caption{Main~$S_i$ and total~$ST_i$ sensitivities from the bi-fidelity SA for an increasing computational budget. Solid and dotted lines represent the sensitivity indices estimated for the unperturbed and perturbed 0D model, respectively. For comparison, fourth order polynomial chaos estimates of the same sensitivity indices are shown with horizontal dashed lines.}
    \label{fig:Sensitivity_vs_budget}
\end{figure}

Bi-fidelity SA was conducted for seven increasing computational budgets. The number of model evaluations per fidelity level and the corresponding control variate coefficients~$\alpha_k$ for all three QoIs are reported in Table~\ref{tab:ModelAllocation_1D_0D}.
For a given computational budget, model allocations were found to change by using perturbed versus unperturbed 0D models results. 
Number of 1D model evaluations increased by factors of 4.8, 5.3, and 4.5 for the QoIs P$_{\mathrm{sys}}$, PP, and $\Delta r_{\mathrm{max}}$, while 0D model allocation reduced by 18\%, 16\%, and 6\%, respectively. 
Reductions in the correlation~$\rho_{1D/0D}$ were observed equal to 0.0145, 0.0174 and 0.0533, for P$_{\mathrm{sys}}$, PP and $\Delta r_{\mathrm{max}}$, respectively.
Estimated main and total Sobol' indices for each computational budget using both the unperturbed (solid lines) or perturbed (dotted lines) low-fidelity are shown in Figure~\ref{fig:Sensitivity_vs_budget}.
Computational budgets 2000 to 10000 show approximately the same estimated Sobol' indices for the unperturbed low-fidelity, thus these indices are regarded as converged, whereas slower convergence is observed for the perturbed case, as expected.

\begin{table}[!ht]
    \scriptsize
    \centering
    \caption{Model allocations for 1D/0D bi-fidelity SA for all QoIs and an increasing computational budget. The allocation $m_k$ represents the model evaluations of the set $\{y^{(A,s)} \}_{s=1}^N$, while the total number of model evaluations is the sum of $\{y^{(A,s)} \}_{s=1}^N$, $\{y^{(B,s)} \}_{s=1}^N$, and $\{y^{(C_j,s)} \}_{s=1}^N$, thus $(d+2)\,m_k = 5\,m_k$.}
    \begin{tabular}{c|l|c |c|c c |c c |c c| c c}
        \toprule
         & & & & \multicolumn{8}{c}{computational budget} \\
         & QoI & model & $\alpha_k$ & \multicolumn{2}{c}{500} & \multicolumn{2}{c}{2000} & \multicolumn{2}{c}{6000} & \multicolumn{2}{c}{10000} \\
         & & & & $m_k$ & total & $m_k$ & total & $m_k$ & total & $m_k$ & total \\
         \midrule
        \multirow{6}{*}{\rotatebox[origin=c]{90}{unperturbed}} & \multirow{2}{*}{P$_{\text{sys}}$} & 1D & 1 & 4 & 20 & 18 & 90 & 55 & 275 & 92 & 460\\
         & & 0D & 0.9916 & 317 & 1585 & 1271 & 6355 & 3814 & 19710 & 6356 & 31780\\
         \cmidrule{2-12}
        & \multirow{2}{*}{PP} & 1D & 1 & 3 & 15 & 12 & 60 & 37 & 185 & 61 & 305\\
         & & 0D & 1.0114 & 323 & 1615 & 1292 & 6460 & 3876 & 19380 & 6460 & 32300 \\
         \cmidrule{2-12}
        & \multirow{2}{*}{$\Delta r_{\text{max}}$} & 1D & 1 & 2 & 10 & 6 & 30 & 19 & 95 & 33 & 165 \\
         & & 0D & 0.9598 & 327 & 1635 & 1311 & 6555 & 3933 & 19665  & 6555 & 32775 \\
        \midrule

        \multirow{6}{*}{\rotatebox[origin=c]{90}{perturbed}} & \multirow{2}{*}{P$_{\text{sys}}$} & 1D & 1 & 21 & 105 & 87 & 435 & 262 & 1310  & 437 & 2185\\
         & & 0D & 0.8611 & 260 & 1300 & 1041 & 5205 & 3124 & 15620 & 5208 & 26040\\
         \cmidrule{2-12}
        & \multirow{2}{*}{PP} & 1D & 1 & 16 & 80 & 65 & 325 & 196 & 980 & 326 & 1630\\
         & & 0D & 0.9769 &  278 & 1390 & 1115 & 5575 & 3346 & 16730 & 5576 & 27880 \\
         \cmidrule{2-12}
        & \multirow{2}{*}{$\Delta r_{\text{max}}$} & 1D & 1 & 7 & 35 & 29 & 145 & 88 & 440 & 148 & 740 \\
         & & 0D & 0.9105 & 308 & 1540  & 1234 & 6170 & 3703 & 18515 & 6172 & 30860\\
         \bottomrule
    \end{tabular}
    \label{tab:ModelAllocation_1D_0D}
\end{table}

All QoIs are most sensitive to variations in the radius~$r$, with lower and equal sensitivity to the wall thickness~$h$ and the elastic modulus~$E$.
Additionally, main and total Sobol' indices are approximately the same, indicating lack of interaction among the random inputs.
For comparison, PC expansion of order four on the 1D model was computed and summarized in Table~\ref{tab:Sensitivities_PC_1D}. Higher polynomial orders were also tested, but the Sobol' indices were already converged for the fourth order.
The bi-fidelity estimates are in good agreement with the PC expansion results.

\begin{figure}[!ht]
    \centering
    \includegraphics[width=\textwidth]{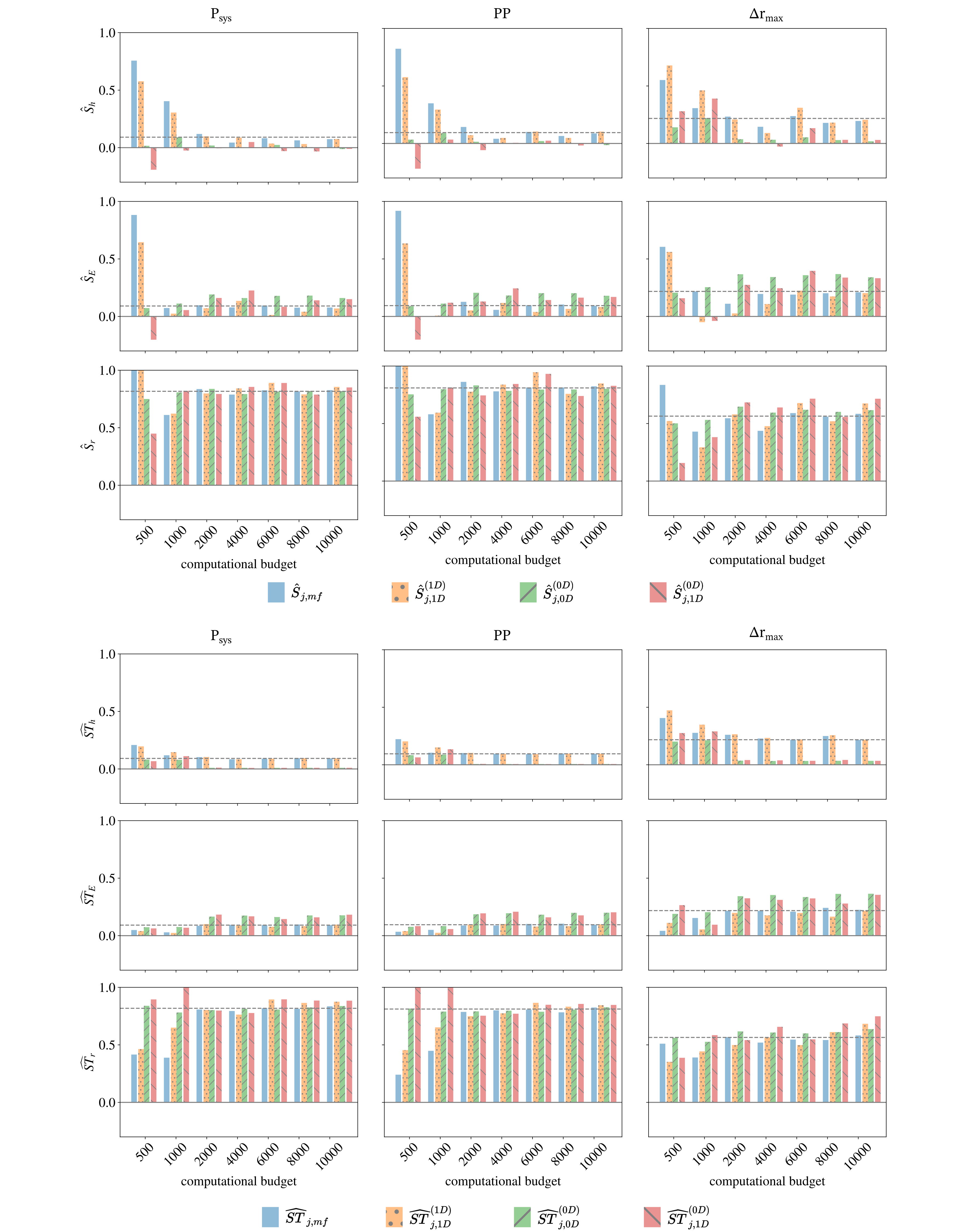}
    \caption{The upper panels show the perturbed bi-fidelity estimates for the main Sobol' sensitivity indices $\hat{S}_{j,mf}$ for all QoIs, as well as the intermediate sensitivity results of $\hat{S}_{j,1D}^{1D}$, $\hat{S}_{j,0D}^{0D}$, and $\hat{S}_{j,1D}^{0D}$. 
    The lower panels are the analogs for the total sensitivity indices $\widehat{ST}_{j,mf}$, $\widehat{ST}_{j,1D}^{1D}$, $\widehat{ST}_{j,0D}^{0D}$, and $\widehat{ST}_{j,1D}^{0D}$.}
    \label{fig:Intermediate_1D_0D_perturbed}
\end{figure}

The estimated components of the bi-fidelity estimator for both the main and total Sobol' indices are illustrated in the Appendix in Figure~\ref{fig:Intermediate_1D_0D} for unperturbed 0D model results, and in Figure~\ref{fig:Intermediate_1D_0D_perturbed} for the perturbed case. These include the final bi-fidelity estimator  $\hat{S}_{j,mf}$ and the two terms $\hat{S}_{j,m_k}^{k}$ and $\hat{S}_{j,m_k-1}^{k}$. 
The multifidelity estimate $\hat{S}_{j,mf}$ is the result of Equations~\eqref{eq:MF_Total_Sobol_Estimator} and~\eqref{eq:MF_Sobol_Estimators}, explicitly $\hat{S}_{j,mf}$ (blue column) equals $\hat{S}_{j,\mathrm{1D}}^{\mathrm{1D}}$ (yellow column) plus the $\alpha_2$-weighted difference of $\hat{S}_{j,\mathrm{0D}}^{\mathrm{0D}}$ (green column) and $\hat{S}_{j,\mathrm{1D}}^{\mathrm{0D}}$(red column).
\begin{table}
    \caption{Sobol' sensitivity indices estimated for the 1D model with fourth order PC expansion.}
    \label{tab:Sensitivities_PC_1D}
    \scriptsize
    \centering
    \begin{subtable}[t]{0.48\textwidth}
        \centering
        \begin{tabular}{l l|c | c|c}
        \toprule
        & & \multicolumn{1}{c}{P$_{\mathrm{sys}}$} & \multicolumn{1}{c}{PP} & \multicolumn{1}{c}{$\Delta r_{\mathrm{max}}$}\\
        \midrule
        \multirow{3}{*}{$\hat{S}_i$} & $h$  &  0.09 & 0.09 & 0.22 \\
        & $E$  &  0.09 & 0.09 & 0.22\\
        & $r$  &  0.82 & 0.81 & 0.56 \\
        \bottomrule
        \end{tabular}
    \end{subtable}
    \begin{subtable}[t]{0.48\textwidth}
        \centering
        \begin{tabular}{l l|c | c|c}
        \toprule
        & & \multicolumn{1}{c}{P$_{\mathrm{sys}}$} & \multicolumn{1}{c}{PP} & \multicolumn{1}{c}{$\Delta r_{\mathrm{max}}$}\\
        \midrule
        \multirow{3}{*}{$\widehat{ST}_i$} & $h$  &  0.09 & 0.09 & 0.22\\
        & $E$  &  0.09 & 0.09 & 0.22 \\
        & $r$  &  0.82 & 0.81 & 0.56 \\
        \bottomrule
        \end{tabular}
    \end{subtable}
\end{table}
Note that $\hat{S}_{j,1D}^{1D} \approx \hat{S}_{j,1D}^{0D}$ (yellow and red column) for all QoIs and all sensitivity indices.  

For a small computational budget, the estimates of sensitivity indices for the 1D model seem inaccurate, but compensate with $\hat{S}_{j,1D}^{0D}$ to produce a reasonable accurate multifidelity estimator.
Intermediate total Sobol' indices show less variations between the different fidelity levels than for the main Sobol' indices.  

\begin{figure}[!h]
    \centering
    \includegraphics[width=\textwidth]{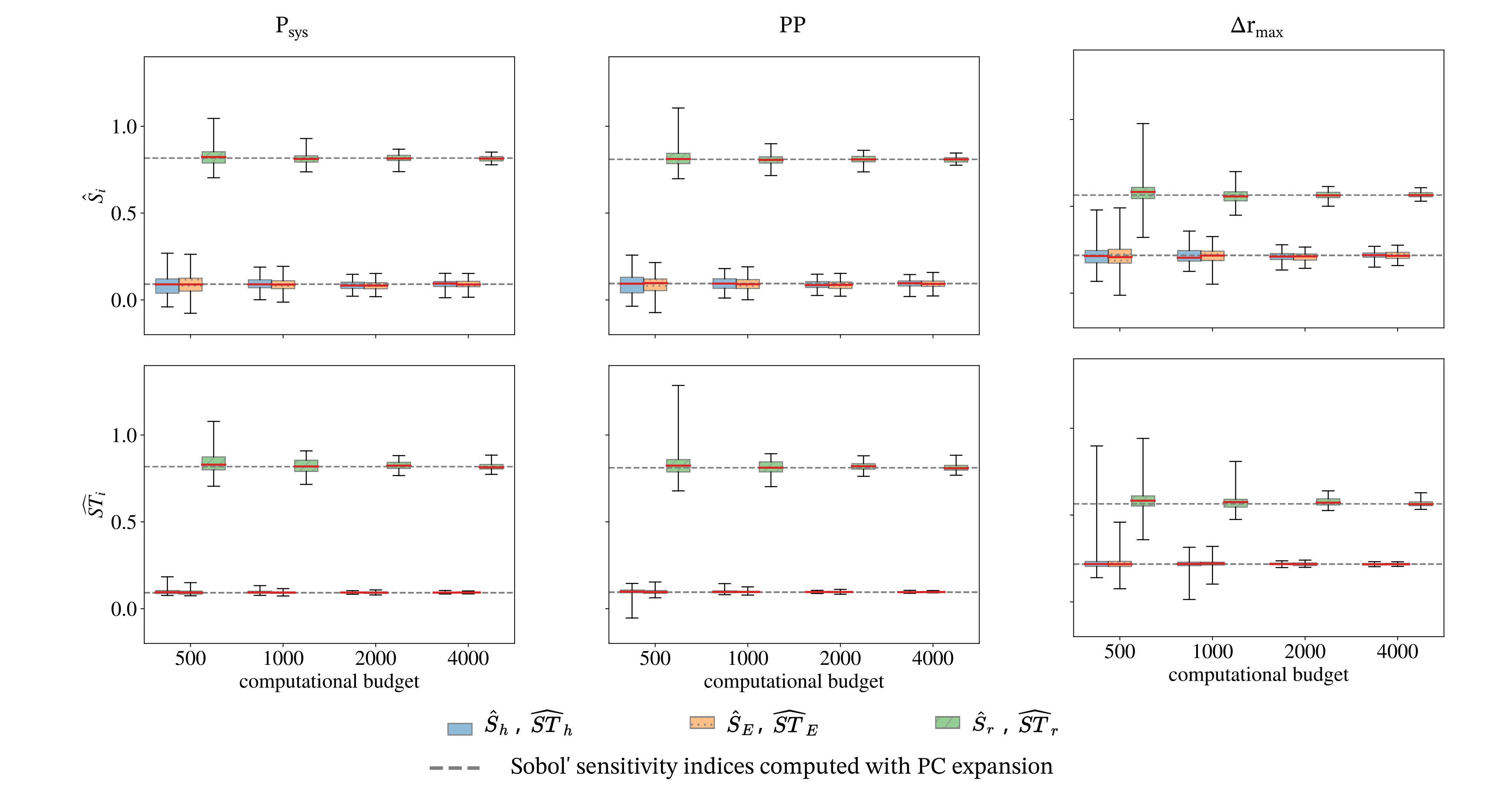}
    \caption{Variance reduction for bi-fidelity MFMC sensitivity indices (100 repetitions) with an increasing computational budget of 500, 1000, 2000, and 4000 simulations. The box indicates the 25$^{\mathrm{th}}$ to 75$^{\mathrm{th}}$ percentile of the sensitivity indices while the whiskers show the minimum and maximum values.}
    \label{fig:Variance_reduction_1D_0D_MFMC}
\end{figure}

\begin{figure}[!h]
    \centering
    \includegraphics[width=\textwidth]{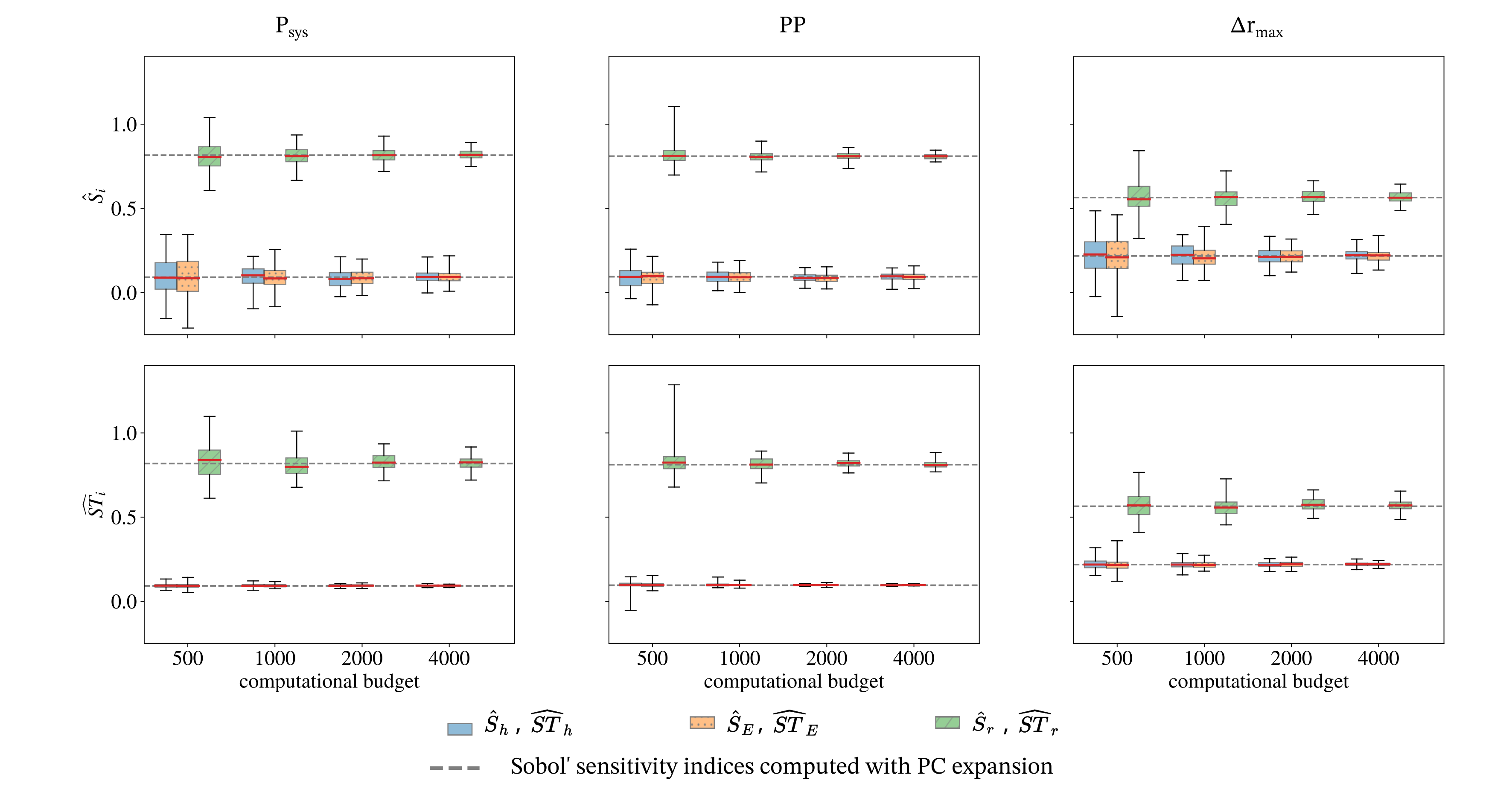}
    \caption{Variance reduction for MC sensitivity indices (100 repetitions) with increasing computational budget of 500, 1000, 2000, and 4000 simulations. The box indicates the 25$^{\mathrm{th}}$ to 75$^{\mathrm{th}}$ percentile of the sensitivity indices, while the whiskers show the minimum and maximum values.}
    \label{fig:Variance_reduction_1D_0D_MC}
\end{figure}

\begin{figure}[!ht]
    \centering
    \includegraphics[width=\textwidth]{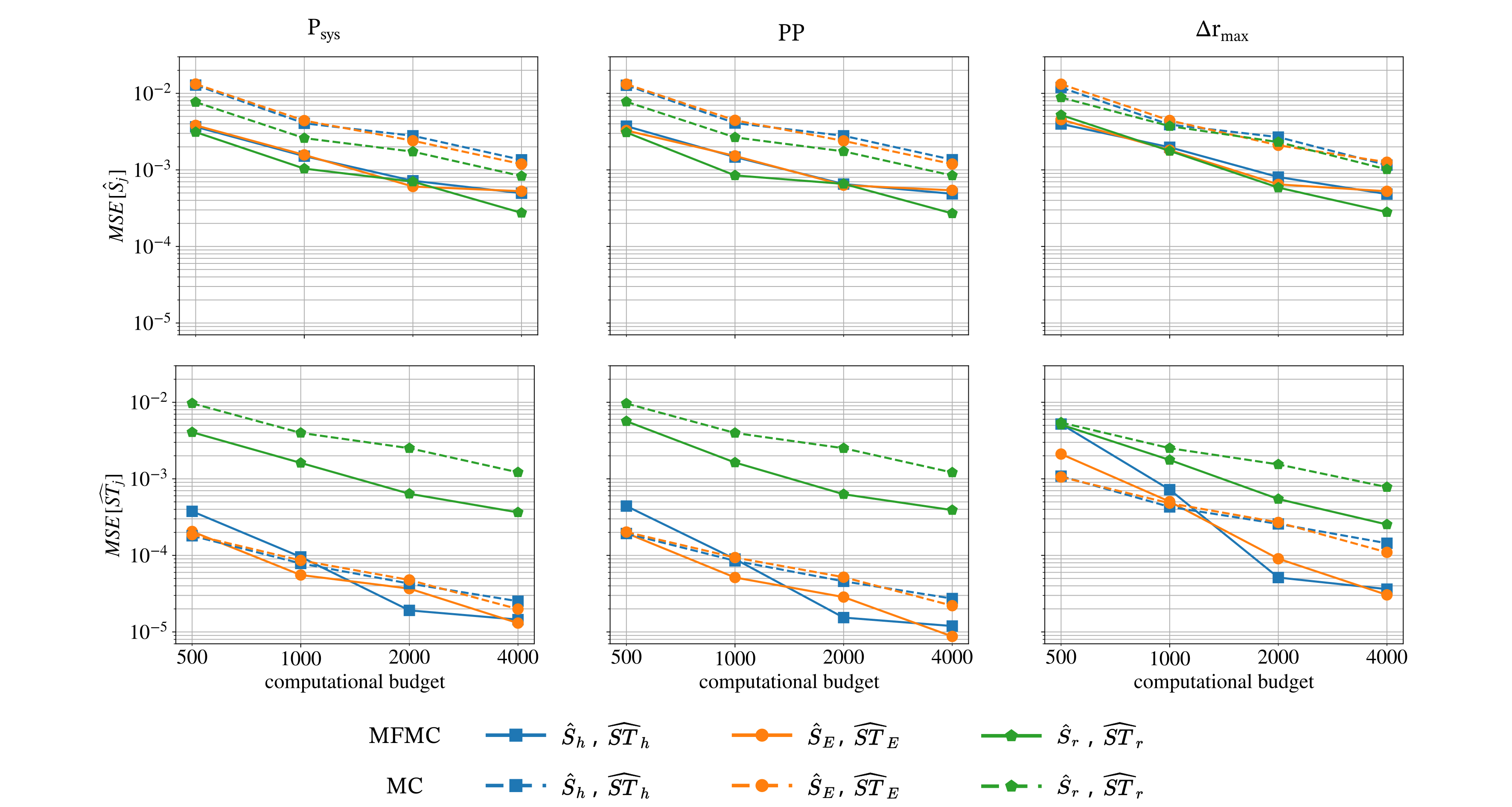}
    \caption{Convergence of bi-fidelity MFMC (solid lines) and Monte Carlo (dashed lines) main (upper row) and total (lower row) Sobol' sensitivity estimates over 100 replications and an increasing computational budget.}
    \label{fig:Convergence_sensitivties_1D_0D}
\end{figure}

Variance reduction in sensitivity indices estimates was investigated with 100 replicates of the unperturbed multifidelity estimator as well as traditional MC (1D model). 
Box plots in Figure~\ref{fig:Variance_reduction_1D_0D_MFMC} and Figure~\ref{fig:Variance_reduction_1D_0D_MC} illustrate the sensitivity indices for the bi-fidelity MFMC estimator and MC, showing higher variability for $\hat{S}_i$ than $\widehat{ST}_i$. 
The variability of MC estimates is always higher than their MFMC counterpart.
Mean and standard deviation of the sensitivity indices computed from replicates are presented in Table~\ref{tab:MFMC_Variations_S_ST} and~\ref{tab:MC_Variations_S_ST} in the Appendix for both methods. 
The mean squared error (MSE, with respect to the mean estimate from the repetitions) for all QoIs, main and total Sobol' indices is shown in Figure~\ref{fig:Convergence_sensitivties_1D_0D}.
For all QoIs, the MSE of $\hat{S}_i$ from MFMC was lower than the one obtained with MC for the same computational budget. This was also the result for $\widehat{ST}_r$, but the MSE of $\widehat{ST}_h$ and $\widehat{ST}_E$ was approximately the same regardless of the QoI and the method they were estimated with. 

Mean and standard deviations of the mean~$\hat\mu$ and variance estimates~$\hat V$ are reported in Table~\ref{tab:Variations_mean_std_QoIs}. 
Regardless of the computational budget and the estimation method, the mean estimates of the mean~$\mathbb{E}[\hat\mu]$ were constant for all QoIs. 
The standard deviation of the mean estimate~$\sigma[\hat\mu]$ decreased for an increasing computational budget, and was smaller for MFMC than for MC. 
Mean estimates of the variance~$\mathbb{E}[\hat{V}]$ and its standard deviation~$\sigma [\hat V]$ showed the same pattern with a constant mean, decreasing standard deviation with increasing computational budget, and lower variation for MFMC in~$\sigma [\hat V]$ than for MC.

\FloatBarrier

\begin{table}[!ht]
    \centering
    \scriptsize
    \caption{Mean and standard deviations of the mean and variance estimates of the QoIs for an increasing computational budget, computed using 100 realizations of the bi-fidelity MFMC estimator ($\hat{\mu}_{mf}, \hat{V}_{mf}$) and MC approach ($\hat{\mu}_{MC}, \hat{V}_{MC}$).}
    \label{tab:Variations_mean_std_QoIs}
    \begin{tabular}{c|c|cc|cc|cc|cc|cc|cc}
    \toprule
    \multirow{3}{*}{method}& \multirow{3}{*}{budget}& \multicolumn{6}{c|}{$\hat \mu$} & \multicolumn{6}{c}{$\hat V$} \\
    \cmidrule{3-14}
     & & \multicolumn{2}{c}{$P_{sys}$ [mmHg]} & \multicolumn{2}{c}{PP} [mmHg]& \multicolumn{2}{c|}{$\Delta r_{\mathrm{max}}$ [mm]} & \multicolumn{2}{c}{$P_{sys}$ [mmHg$^2$]} & \multicolumn{2}{c}{PP [mmHg$^2$]} & \multicolumn{2}{c}{$\Delta r_{\mathrm{max}}$ [mm$^2$]} \\
    & & $\mathbb{E}[\hat{\mu}]$ & $\sigma[\hat{\mu}]$ & $\mathbb{E}[\hat{\mu}]$ & $\sigma[\hat{\mu}]$ & $\mathbb{E}[\hat{\mu}]$ & $\sigma[\hat{\mu}]$ & $\mathbb{E}[\hat{V}]$ & $\sigma[\hat{V}]$ & $\mathbb{E}[\hat{V}]$ & $\sigma[\hat{V}]$ & $\mathbb{E}[\hat{V}]$ & $\sigma[\hat{V}]$\\
    \midrule
    \multirow{4}{*}{MFMC} & 500 & 129.600 & 0.072 & 51.039 & 0.112 & 0.162 & 0.544 & 3.273 & 0.158 & 7.910 & 0.365 & 0.000 & 9.328\\
    & 1000 & 129.604 & 0.057 & 51.046 & 0.087 & 0.162 & 0.415 & 3.271 & 0.104 & 7.907 & 0.244 & 0.000 & 7.121\\
    & 2000 & 129.596 & 0.036 & 51.031 & 0.054 & 0.162 & 0.265 & 3.276 & 0.072 & 7.925 & 0.167 & 0.000 & 4.705\\
    & 4000 & 129.596 & 0.025 & 51.033 & 0.038 & 0.162 & 0.197 & 3.277 & 0.048 & 7.927 & 0.118 & 0.000 & 3.623\\
    \midrule
        \multirow{4}{*}{MC} & 500 & 129.596 & 0.129 & 51.031 & 0.202 & 0.162 & 1.031 & 3.286 & 0.258 & 7.950 & 0.633 & 0.000 & 17.862\\
        & 1000 & 129.599 & 0.091 & 51.036 & 0.141 & 0.162 & 0.745 & 3.264 & 0.198 & 7.898 & 0.485 & 0.000 & 12.631\\
        & 2000 & 129.595 & 0.057 & 51.030 & 0.090 & 0.162 & 0.461 & 3.249 & 0.127 & 7.859 & 0.311 & 0.000 & 8.828\\
        & 4000 & 129.600 & 0.043 & 51.037 & 0.067 & 0.162 & 0.340 & 3.260 & 0.085 & 7.885 & 0.209 & 0.000 & 6.226\\
    \bottomrule
    \end{tabular}
\end{table}

\begin{figure}[!ht]
    \centering
    \includegraphics[width=\textwidth]{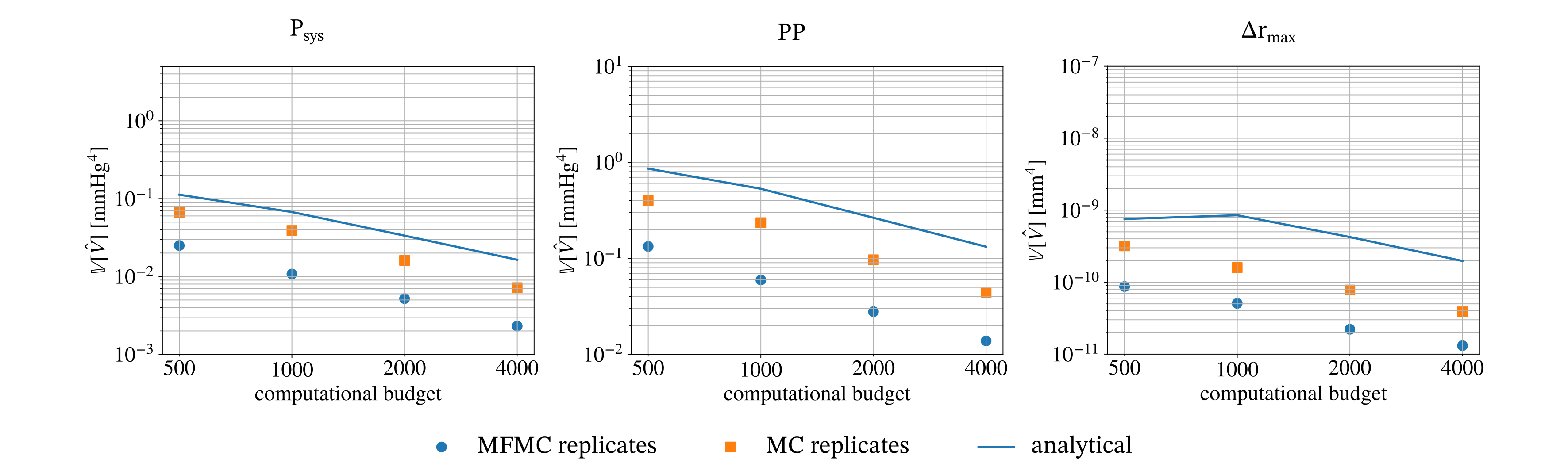}
    \caption{Convergence of bi-fidelity variance estimates. The line represents the theoretical values from Equation~\ref{eq:analytical_Variance} while the estimates for computational budgets 500, 1000, 2000, and 4000 are computed over 100 repetitions.}
    \label{fig:MSE_variance_estimator_1D_0D}
\end{figure}
 
Sample variance computed through repetitions for the computational budgets 500-4000 are shown in Figure~\ref{fig:MSE_variance_estimator_1D_0D}.
The continuous line represents the analytical variance of the multi-fidelity variance estimator from Equation~\eqref{eq:analytical_Variance}, while the MFMC and MC replicates are indicated using disks and squares, respectively. 
The estimated variance of the multifidelity variance estimator is lower than predicted analytically for both MFMC and MC.

\subsection{Results of 3D/1D/0D tri-fidelity SA}

\begin{table}[!ht]
    \scriptsize
    \centering
    \caption{Model allocations for 3D/1D/0D tri-fidelity SA for all QoIs at increasing computational budgets. The allocation $m_k$ represents the model evaluations of the set $\{y^{(A,s)} \}_{s=1}^N$, while the total number of model evaluations is the sum of $\{y^{(A,s)} \}_{s=1}^N$, $\{y^{(B,s)} \}_{s=1}^N$, and $\{y^{(C_j,s)} \}_{s=1}^N$, thus $(d+2)\,m_k = 5\,m_k$. The last column indicates the difference~$\Delta$ in model allocation with respect to 150 pilot runs.}
    \begin{tabular}{l |c|l |c|c c |c c| c c| c c | c}
        \toprule
         & & & &  \multicolumn{8}{c}{computational budget} \\
         QoI & \# pilot runs& model& $\alpha_k$ & \multicolumn{2}{c}{20} & \multicolumn{2}{c}{30} & \multicolumn{2}{c}{40} & \multicolumn{2}{c}{50} & \\
         & & & & $m_k$ & total & $m_k$ & total & $m_k$ & total & $m_k$ & total & $\Delta$ [\%]\\
        \midrule
         \multirow{9}{*}{P$_{\text{sys}}$} & \multirow{3}{*}{10} & 3D-FSI model & 1 & 3 & 15 & 5 & 25 & 7 & 35 & 9 & 45 & 0\\
        & & 1D-model & 1.0574 & 405 & 2025 & 608 & 3040 & 811 & 4055 & 1014 & 5070 & +23.3 \\
        & & 0D-model & 1.1112 & 3657 & 18285 & 5486 & 27430 & 7315 & 36575 & 9143 & 45715 & +2.9 \\
        \cmidrule{2-13}
        & \multirow{3}{*}{20} & 3D-FSI model & 1 & 3 & 15 & 5 & 25 & 7 & 35 & 9 & 45 & 0 \\
        & & 1D-model & 1.0693 & 263 & 1315 & 394 & 1970 & 526 & 2630 & 657 & 3285 & -20.1 \\
        & & 0D-model & 1.1022 & 5290 & 26450 & 7935 & 39675 & 10580 & 52900 & 13225 & 66125 & +48.8 \\
         \cmidrule{2-13}
         & \multirow{3}{*}{150} & 3D-FSI model & 1 & 3 & 15 & 5 & 25 & 7 & 35 & 9 & 45 & - \\
         & & 1D-model & 1.0524 & 329 & 1645 & 493 & 2465 & 658 & 3290 & 822 & 4110 & - \\
         & & 0D-model & 1.0449 & 3554 & 17770 & 5331 & 26655 & 7108 & 35540 & 8885 & 44425 & - \\
         
         \midrule
        \multirow{9}{*}{PP} & \multirow{3}{*}{10} & 3D-FSI model & 1 & 3 & 15 & 5 & 25 & 7 & 35 & 9 & 45 & 0\\
        & & 1D-model & 1.0302 & 401 & 2005 & 602 & 3010 & 802 & 4010 & 1003 & 5015 & +11.1 \\
        & & 0D-model & 1.0701 & 3484 & 17420 & 5227 & 26135 & 6969 & 34845 & 8712 & 43560 & +1.4 \\
        \cmidrule{2-13}
        & \multirow{3}{*}{20} & 3D-FSI model & 1 & 3 & 15 & 5 & 25 & 7 & 35 & 9 & 45 & 0 \\
        & & 1D-model & 1.0498 & 167 & 835 & 251 & 1255 & 334 & 1670 & 418 & 2090 & -53.8 \\
        & & 0D-model & 1.1049 & 5388 & 26940 & 8082 & 40410 & 10777 & 53885 & 13471 & 67355 & +56.8 \\
         \cmidrule{2-13}
        & \multirow{3}{*}{150} & 3D-FSI model & 1 & 3 & 15 & 5 & 25 & 7 & 35 & 9 & 45 & - \\
        & & 1D-model & 1.0278 & 361 & 1805 & 542 & 2710 & 723 & 3615 & 904 & 4520 & - \\
        & & 0D-model & 1.0301 & 3436 & 17180 & 5155 & 25775 & 6873 & 34365 & 8591 & 42955 & - \\
        
        \midrule
         \multirow{9}{*}{$\Delta r_{\text{max}}$} & \multirow{3}{*}{10} & 3D-FSI model & 1 & 3 & 15 & 5 & 25 & 7 & 35 & 9 & 45 & 0\\
        & & 1D-model & 1.1391 & 583 & 2915 & 874 & 4370 & 1166 & 5830 & 1457 & 7285 & +50.6\\
        & & 0D-model & 1.1252 & 1897 & 9485 & 2846 & 14230 & 3795 & 18975 & 4744 & 23720 & +0.3\\
        \cmidrule{2-13}

        & \multirow{3}{*}{20} & 3D-FSI model & 1 & 3 & 15 & 5 & 25 & 7 & 35 & 9 & 45 & 0 \\
        & & 1D-model & 1.1386 & 173 & 865 & 259 & 1295 & 346 & 1730 & 433 & 2165 & -53.3 \\
        & & 0D-model & 1.2782 & 2720 & 13600 & 4081 & 20405 & 5441 & 27205 & 6802 & 34010 & +43.9\\
         
         \cmidrule{2-13}
         & \multirow{3}{*}{150} & 3D-FSI model & 1 & 3 & 15 & 5 & 25 & 7 & 35 & 9 & 45 & - \\
         & & 1D-model & 1.0879 & 387 & 1935 & 581 & 2905 & 775 & 3875 & 969 & 4845 & - \\
         & & 0D-model & 1.0540 & 1891 & 9455 & 2836 & 14180 & 3782 & 18910 & 4728 & 23640 & - \\
         \bottomrule
    \end{tabular}
    \label{tab:ModelAllocation_3D_1D_0D}
\end{table}

\begin{figure}[!ht]
    \centering
    \includegraphics[width=\textwidth]{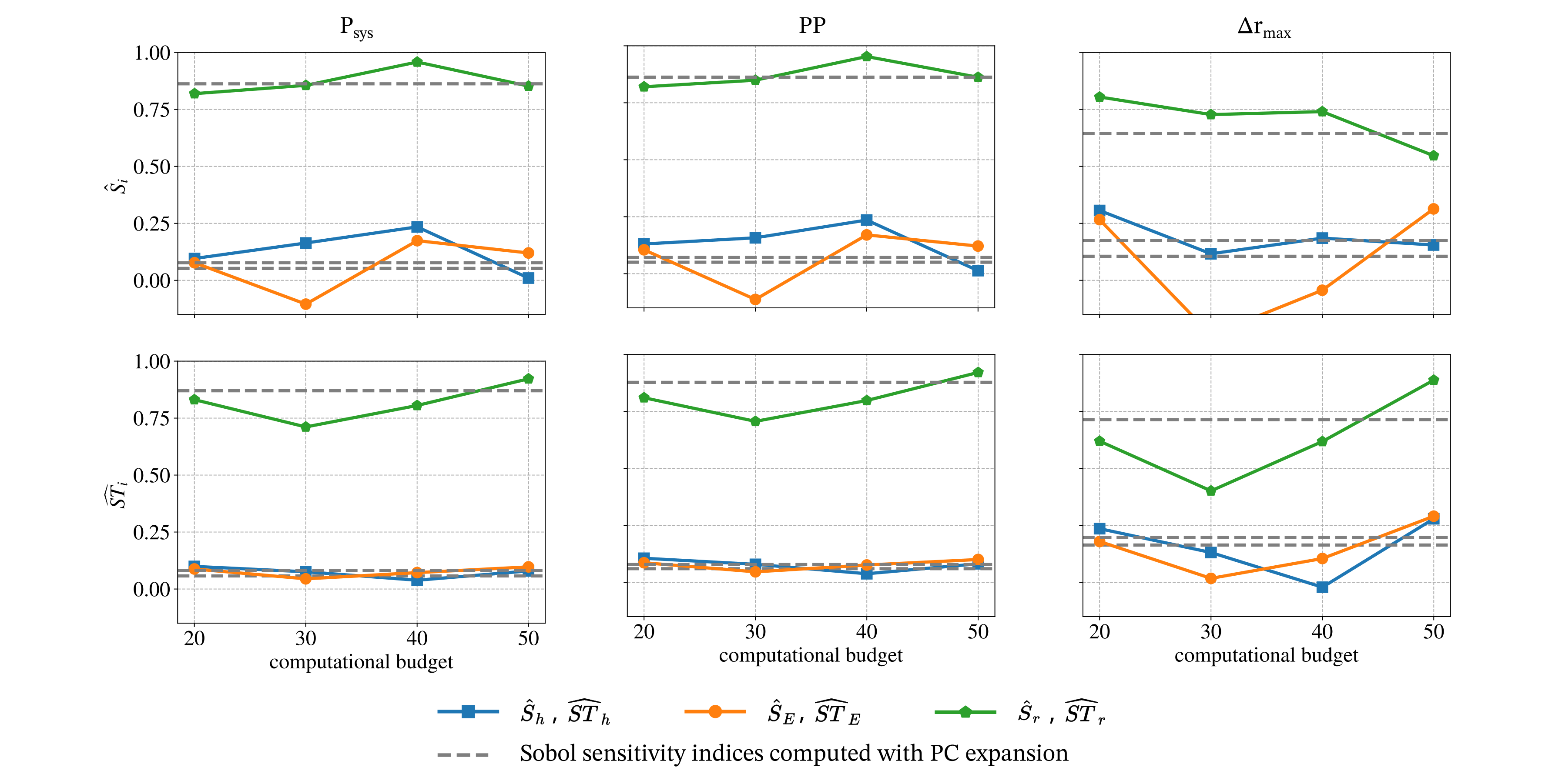}
    \caption{MFMC estimates of main ($\hat{S}_i$) and total ($\widehat{ST}_i$) Sobol' indices for increasing computational budgets. Indices computed from order-three PC are represented with horizontal dashed lines.}
    \label{fig:Sensitivity_vs_budget_3_levels}
\end{figure}

\begin{table}[!ht]
    \scriptsize
    \centering
    \caption{Sobol' sensitivity indices estimated for 3D-FSI model with PC expansion.}
    \label{tab:Sensitivities_PC_3D}
    \begin{tabular}{l l|cccc | cccc|cccc}
    \toprule
    & & \multicolumn{4}{c}{P$_{\mathrm{sys}}$} & \multicolumn{4}{c}{PP} & \multicolumn{4}{c}{$\Delta r_{\mathrm{max}}$}\\
    \midrule
    \multicolumn{2}{c|}{PC order} & 1 & 2 & 3 & 4 & 1 & 2 & 3 & 4 & 1 & 2 & 3 & 4 \\
    \midrule
    \multirow{3}{*}{$\hat{S}_i$} & $h$  & 0.05 & 0.08 & 0.05 & 0.09 & 0.06 & 0.08 & 0.05 & 0.09 & 0.14 & 0.15 & 0.11 & 0.25 \\
    & $E$  & 0.10 & 0.09 & 0.08 & 0.09 & 0.09 & 0.09 & 0.07 & 0.09 & 0.25 & 0.17 & 0.17 & 0.16 \\
    & $r$  & 0.85 & 0.83 & 0.86 & 0.80 & 0.85 & 0.83 & 0.86 & 0.81 & 0.61 & 0.65 & 0.64 & 0.43 \\
    \midrule
    \multirow{3}{*}{$\widehat{ST}_i$} & $h$  &  0.05 & 0.08 & 0.06 & 0.11 & 0.06 & 0.08 & 0.06 & 0.10 & 0.14 & 0.18 & 0.16 & 0.37 \\
    & $E$  & 0.10 & 0.09 & 0.08 & 0.10 & 0.09 & 0.09 & 0.08 & 0.09 & 0.25 & 0.17 & 0.20 & 0.24 \\
    & $r$  & 0.85 & 0.84 & 0.87 & 0.81 & 0.85 & 0.84 & 0.88 & 0.82 & 0.61 & 0.68 & 0.71 & 0.58 \\
    \toprule
\end{tabular}
\end{table}

Four increasing computational budgets were tested for the tri-fidelity MFMC SA. The number of model allocations per computational budget and fidelity level as well as the corresponding control variate coefficients~$\alpha_k$ are presented in Table~\ref{tab:ModelAllocation_3D_1D_0D}. 
The same 3D-FSI model allocations were obtained independent of the number of pilot runs. 
Also, a decrease from 150 to 10 pilot run simulations led to an increase in the 1D model allocations for all QoIs, while the number of 0D evaluations stayed approximately constant. Highest increase showed $\Delta r_{\mathrm{max}}$ with 50.5\% and lowest increase was observed for PP with 11.1\%. Computing model allocation based on model statistics of only 20 pilot runs, the number of model evaluations for 1D decreased between 20.1\% and 53.3\% and the number of 0D model evaluations increased by 43.9\% to 56.8\% compared to estimated model statistics from 150 pilot runs. Smallest changes in model allocations were in P$_{\mathrm{sys}}$ and largest changes in PP.

MFMC estimates for the main and total sensitivity indices were compared to the high-fidelity fourth order PC in Figure~\ref{fig:Sensitivity_vs_budget_3_levels}. 
The importance ranks suggested by the sensitivity indices for each QoI are the same as in the bi-fidelity approximation, but the values differ slightly.
For the polynomial orders~1-4, the main and total Sobol' indices are presented in Table~\ref{tab:Sensitivities_PC_3D}, with slight variations between orders.
Estimated $\hat{S_i}$ and $\widehat{ST}_i$ from MFMC are close to PC indices for P$_{\text{sys}}$ and PP, while a larger discrepancy is observed for $\Delta r_{max}$.  

\begin{figure}[!ht]
    \centering
    \includegraphics[width=\textwidth]{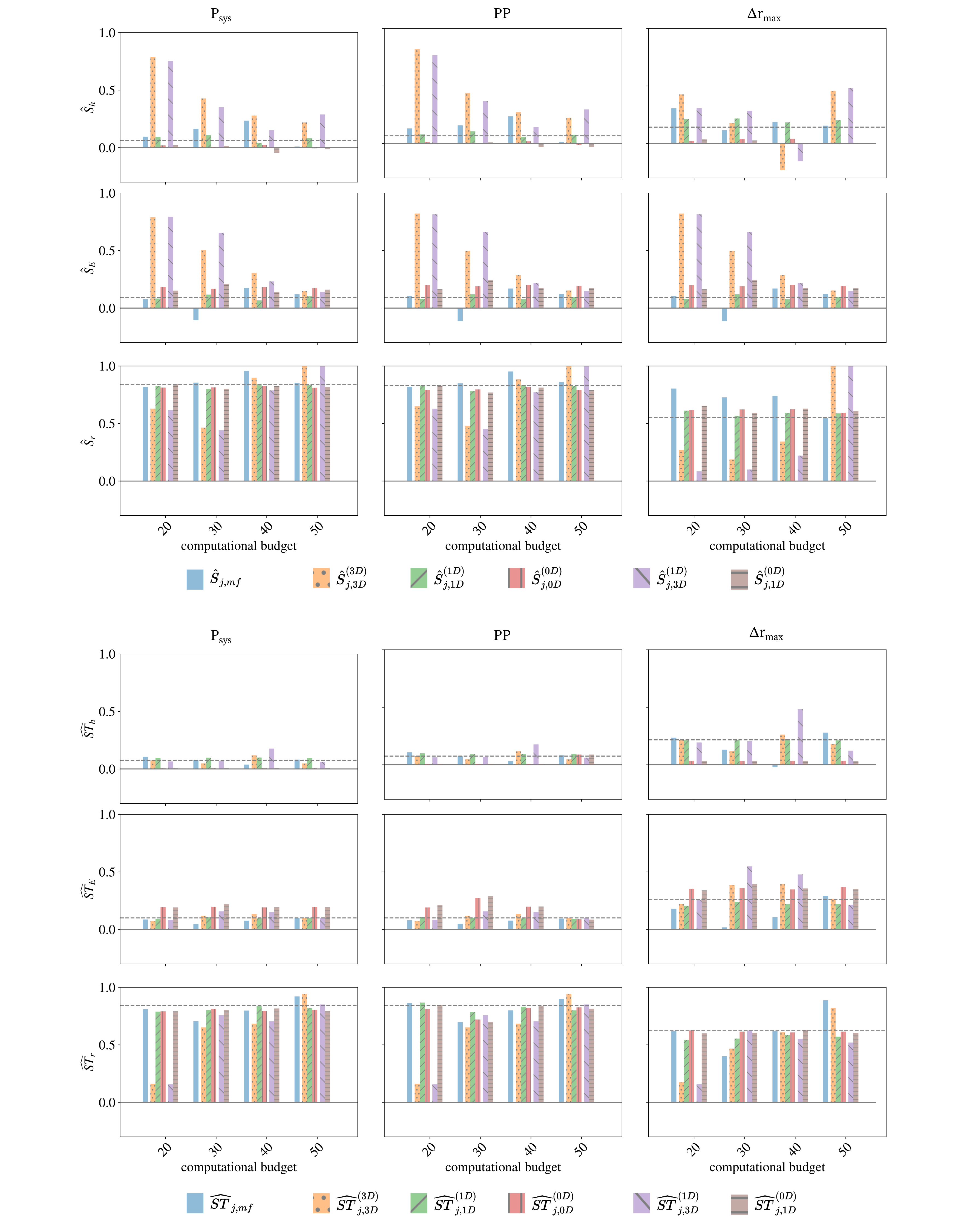}
    \caption{The upper panels show the tri-fidelity estimates for the main indices $\hat{S}_{j,mf}$ for the QoIs P$_{\mathrm{sys}}$, PP, and $\Delta r_{\mathrm{max}}$. Intermediate sensitivity results $\hat{S}_{j,1D}^{1D}$, $\hat{S}_{j,0D}^{0D}$, and $\hat{S}_{j,1D}^{0D}$ are also displayed. The lower panels contain both the estimate $\widehat{ST}_{j,mf}$, and the three terms $\widehat{ST}_{j,1D}^{1D}$, $\widehat{ST}_{j,0D}^{0D}$, and $\widehat{ST}_{j,1D}^{0D}$.}
    \label{fig:Intermediate_3D_1D_0D}
\end{figure}

The terms contributing to the tri-fidelity MFMC estimates are displayed in Figure~\ref{fig:Intermediate_3D_1D_0D}. 
For each computational budget, five bars are shown, the first representing the MFMC estimates $\hat{S}_{j,mf}$ and $\widehat{ST}_{j_mf}$, respectively, the second $\hat{S}_{j, m_k}^{k}$ and $\widehat{ST}^{k}_{j,m_k}$, and the third representing $\hat{S}_{j, m_{k-1}}^{k}$ and $\widehat{ST}_{j,m_{k-1}}^{k}$.
Here the bad approximation of the MFMC indices becomes apparent since $\hat{S}_{j, 3D}^{3D} \neq \hat{S}_{j,3D}^{1D}$ and $\hat{S}_{j, 1D}^{1D} \neq \hat{S}_{j,1D}^{0D}$. 
For cases where instead $\hat{S}_{j, 3D}^{3D} \approx \hat{S}_{j,3D}^{1D}$ and $\hat{S}_{j, 1D}^{1D} \approx \hat{S}_{j,1D}^{0D}$, then the MFMC and PC estimates become close.

\begin{figure}[!ht]
    \centering
    \includegraphics[width=\textwidth]{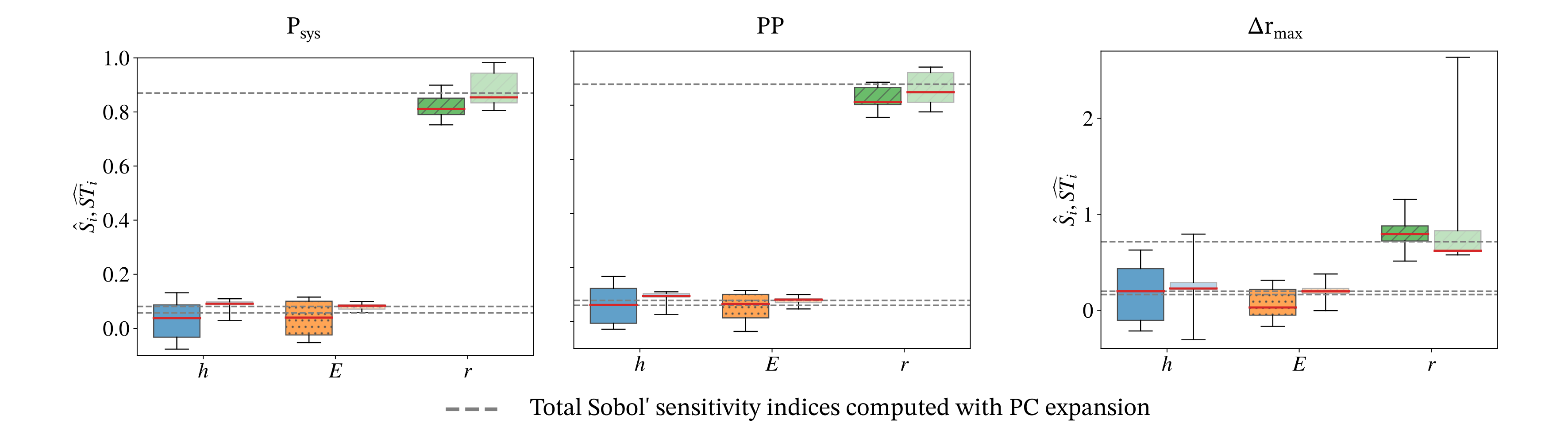}
    \caption{Tri-fidelity MFMC estimates of sensitivity indices from 6 repetitions, each with budget 20. Two box plots are shown for each input, representing main (left) and total (right) sensitivity indices. Each box indicates the 25th to 75th percentile of the sensitivity index, while the whiskers show the minimum and maximum values. Order-three PC estimates are also plotted using horizontal lines. Note the different scale of the y-axis for $\Delta r_{\text{max}}$.}
    \label{fig:Boxplot_3D_mf_sensitivities}
\end{figure}

\begin{figure}[!ht]
    \centering
    \includegraphics[width=\textwidth]{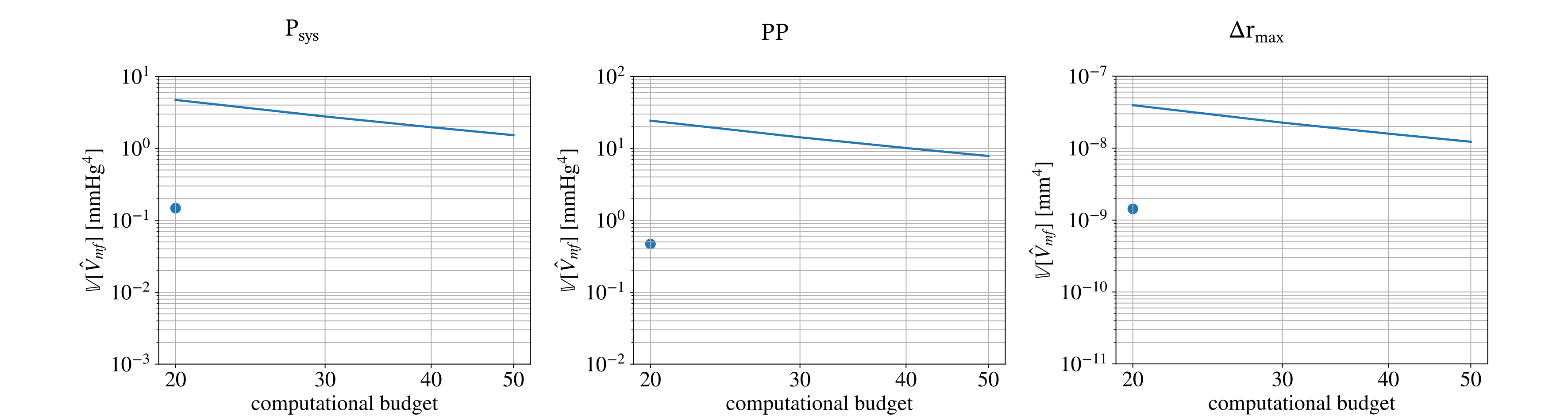}
    \caption{Convergence of tri-fidelity MFMC variance estimates. The line represents the analytical MSE following Equation~\eqref{eq:analytical_Variance} and the variance estimate for the computational budget of~20 computed with 6 replicates is indicated with a disk.} 
    \label{fig:MSE_variance_estimator_3D_1D_0D}
\end{figure}

All available 3D-FSI simulations were used to investigate variance reduction through 6 repeated evaluations of the estimator resulting from the lowest computational budget. 
%
%
Variations in the main and total Sobol' sensitivity indices from tri-fidelity estimators for these 6~repetitions are presented in Figure~\ref{fig:Boxplot_3D_mf_sensitivities}.
MFMC and PC estimates are found close for P$_{\mathrm{sys}}$ and PP, with differences for $\widehat{ST}_i$ generally smaller than for $\hat{S}_i$. 
For $\Delta r_{\mathrm{max}}$, mean estimates of $\hat{S}_i$ and $\widehat{ST}_i$ are also comparable to the results from PC, but their variations are larger than for the other QoIs. 
There are also significant outliers in the Sobol' sensitivity indices for $\Delta r_{\mathrm{max}}$. 
Computed mean values and standard deviations from these repeated evaluations are reported in Table~\ref{tab:triFidelity_Variations_S_ST} in the Appendix.

Figure~\ref{fig:MSE_variance_estimator_3D_1D_0D} shows that the variance resulting from the repeated evaluations of the computational budget~20 MFMC estimates is lower than the analytical expression in Equation~\ref{eq:analytical_Variance}, this latter computed from the values reported in Table~\ref{tab:ModelStatistics_3D_1D_0D}.
Greater variance reduction is achieved by the bi-fidelity estimator with respect to the tri-fidelity results for the tested computational budgets.

\subsection{Comparison of computational expenses}\label{sec:comp_cost}
\begin{table}[!ht]
    \scriptsize
    \centering
    \caption{Computational cost for bi-fidelity SA comparing MFMC, MC, and PC expansion, where the equivalent polynomial order and the number of 1D model evaluations are also reported.}
    \begin{tabular}{l l |r r r r r r r}
        \toprule
         \multicolumn{2}{l}{SA method} & & \multicolumn{6}{c}{computational budget} \\
           & & 500 & 1000 & 2000 & 4000 & 6000 & 8000 & 10000 \\
        \midrule
         MFMC & & 247\,s & 493\,s & 996\,s & 2003\,s &  2980\,s & 3948\,s & 4958\,s \\
         MFMC & perturbed & 359\,s & 659\,s & 1343\,s & 2666\,s &  4028\,s & 5344\,s & 6716\,s \\
         MC & & 653\,s & 1315\,s & 2630\,s & 5260\,s & 7776\,s & 10400\,s & 12970\,s \\
         \midrule
         \multirow{3}{*}{PC} & order & 2 & 3 & 4 & 6 & 7 & 8 & 9 \\
         & \# evaluations & 20 & 45 & 90 & 185 & 275 & 370 & 460\\
         & run time & 28\,s & 30\,s & 46\,s & 105\,s & 135\,s & 195\,s & 265\,s \\
         \bottomrule
    \end{tabular}
    \label{tab:Computational_expenses_1D_0D}
\end{table}

\begin{table}[!ht]
    \scriptsize
    \centering
    \caption{Computational costs for tri-fidelity SA comparing MFMC and PC expansion, where additionally the equivalent polynomial order and number of 3D model evaluations is given.}
    \begin{tabular}{l l|r r r r }
        \toprule
         & \multicolumn{4}{c}{computational budget} \\
         & & 20 & 30 & 40 & 50 \\
        \midrule
         MFMC & & 1363104\,s & 3600952\,s & 38790300\,s & 5525081\,s \\
         \midrule
         \multirow{3}{*}{PC} & order & 1 &  2 & 3 & 4 \\
         & \# evaluations & 15 & 25 & 35 & 45  \\
         &  run time & 1358100\,s & 3593560\,s & 3870300\,s & 5525081\,s \\ 
         \bottomrule
    \end{tabular}
    \label{tab:Computational_expenses_3D_1D_0D}
\end{table}

Computational costs for MFMC, MC, and PC expansion are presented for the bi-fidelity scenario in Table~\ref{tab:Computational_expenses_1D_0D}. 
To perform a just comparison between PC expansion with the other methods, the PC order was chosen such that the number of model evaluations needed for determining the polynomial coefficients in the PC expansion with an overdetermined system of equations was the same as the number of 1D model evaluations for the MFMC method for a specific computational budget. 
MFMC with the unperturbed 0D model was approximately 2.6~times faster than traditional MC for the same computational budget, while MFMC with the perturbed 0D model was 1.3 times slower compared to the unperturbed MFMC. PC expansion outperformed MFMC by a factor of 10 to 12 for the investigated computational budgets. 

Table~\ref{tab:Computational_expenses_3D_1D_0D} compares the computational cost of tri-fidelity MFMC and PC expansion. As in the bi-fidelity case, the same number of 3D-FSI model evaluations was used for PC coefficient estimation as for MFMC with an equivalent computational budget.

MFMC has always slightly higher computational costs than PC expansion because the costs for MFMC consist of the expenses for the 3D-FSI model and the expenses of the lower fidelity models, while PC expansion has only the expenses of the high-fidelity model.

\section{Discussion and conclusions}\label{sec:discussion}

In this work, we applied the framework of MFMC global sensitivity analysis presented by Qian et al.~\cite{qian_multifidelity_2018} to an idealized CCA model using three model fidelities, i.e. a high-fidelity 3D-FSI model and two low-fidelity 1D and 0D haemodynamic models.
Systolic pressure~P$_{\mathrm{sys}}$, pulse pressure~PP, and maximal radial displacement $\Delta r_{\text{max}}$, were selected as relevant QoIs.
Lower fidelity models were validated against the 3D-FSI model and a mesh convergence study was performed to determine the number of finite elements leading to the best trade off between cost and accuracy.
Result statistics, correlations and optimal allocations were determined through pilot runs with an increasing number of simulations. 
Global sensitivity indices were then evaluated for 1D/0D bi-fidelity and 3D/1D/0D tri-fidelity MFMC estimators and compared with MC and PC expansion.

Pressure waveforms, flow rates, and radial displacements were found to agree well for all model fidelities, with average errors for pressures and displacements in the range 0.5\,\% and 5\,\%, and highest relative error $\epsilon_{\Delta r}^{3D/0D} = 5.75$\,\%.
The agreement between model results at different fidelities is comparable to those found in other multifidelity studies in the literature~\cite{seo_multifidelity_2020, fleeter_multilevel_2020}. 
%

%
%
%

The first two statistical moments for each QoI were computed from the pilot runs for all model fidelities. The mean values for P$_{\text{sys}}$ and PP are consistent with clinical measurements from 40-year old patients~\cite{vriz_comparison_2017, hansen_diameter_1995, raninen_carotid_2002}. 
Mean values of $\Delta r_{\text{max}}$ are comparable with those found in some studies~\cite{raninen_carotid_2002}, but slightly lower than reported in others~\cite{vriz_comparison_2017, hansen_diameter_1995, riley_variation_1997,engelen_reference_2015}. 
%

Correlations between models are slightly higher than in previous multifidelity studies with patient-specific anatomies~\cite{fleeter_multilevel_2020, seo_multifidelity_2020, schiavazzi_multifidelity_2020}, possibly due to an idealized geometry considered in this study. 
Correlations were especially high for 0D-models due to their ability to accurately estimating bulk haemodynamic quantities~\cite{ceserani_lumped-parameter_2023, mantero_coronary_1992, shi_review_2011}, especially in arterial systems with unobstructed geometries. 
%
%
%
However, 0D model results had to be perturbed in order to compute optimal allocations from Equation~\eqref{eq:optimal_allocation}, with negligible effects on the resulting sensitivity indices and computational cost. 
%
%
%
%
%
Estimating tri-fidelity model statistics for a varying number of pilot run simulations showed limited differences in tri-fidelity estimator moments and allocations between 20 and 150 model evaluations. 
%
%

%
%
%
Breaking down the various contributions to the bi-fidelity estimates for $\hat{S}_i$ and $\widehat{ST}_i$ shows that small computational budgets can lead to poor performance in the single-fidelity scenario.
However, this effect is balanced out if $\hat{S}_{j,1D}^{1D} \approx \hat{S}_{j,1D}^{0D}$ ($\widehat{ST}_{j,1D}^{1D} \approx \widehat{ST}_{j,1D}^{0D}$), leading to a reasonable MFMC estimate.   
%
%
%
Convergence of estimators for main Sobol' indices is consistently faster for MFMC than MC. This is also true for $\widehat{ST}_r$, as opposed to $\widehat{ST}_h$ and $\widehat{ST}_E$ where similar convergence was found for MFMC and MC, possibly due to the significant smaller magnitude of such indices ($\widehat{ST}_r = 0.82$ versus $\widehat{ST}_h = \widehat{ST}_E = 0.09$ for P$_{\mathrm{sys}}$). 
%
Tri-fidelity MFMC indices showed good agreement with PC estimates from the 3D-FSI model for most computational budgets, where variability in the former is expected due to the relatively small number of high-fidelity evaluations per computational budget.
A detailed analysis of the terms in the MFMC estimator helps to understand the conditions leading to this difference, namely when $\hat{S}_{j,3D}^{(3D)} \neq \hat{S}_{j,3D}^{(1D)}$ and $\hat{S}_{j,1D}^{(1D)} \neq \hat{S}_{j,1D}^{(0D)}$. 

Computational costs for global SA in an idealized CCA can be reduced with bi- and tri-fidelity MFMC compared to traditional MC. However, PC outperforms MFMC, due to the smooth underlying stochastic response.
The same results, though, cannot be expected for cases with non-idealized flow conditions. 

This study contains several limitations which should be addressed in future work. 
Uniform distributions are assumed for the uncertain parameters with a variation of 10\% around the mean value. 
Future work will investigate more realistic distributions for the random inputs like a Beta or Gaussian distribution, and additional sources of uncertainty, e.g., in the inlet flow, fluid properties, and the Windkessel boundary conditions.
We considered three scalar QoIs, P$_{\mathrm{sys}}$, PP, and $\Delta r_{\mathrm{max}}$, however, in a clinical setting, time-dependent QoIs are often of interest. 
Additional QoIs of clinical relevance (e.g. the time averaged wall shear stress) will also be included in future work, as well as patient specific anatomies characterized by a smaller correlation among the fidelities.
%
%
%
Another simplification consisted of considering a linear elastic material model for the wall mechanics in the lower fidelity models, and a neo-Hookean material for the 3D-FSI model. 
Future work will focus on more realistic constitutive formulations for the vascular tissue.




\subsection*{Acknowledgements}

The research leading to these results is funded by the Norwegian Financial Mechanism 2014-2021 operated by the National Science Center, PL (NCN) within the GRIEG programme under grant\#~UMO-2019/34/H/ST8/00624, project \textit{non-invasivE iN-vivo assessmenT Human aRtery wALls (ENTHRAL, www.enthral.pl)}. 
DES is partially supported by NSF CAREER award \#1942662 and NSF CDS\&E award \#2104831. We thank Martin Pfaller, Vijaj Vedula, and Chi Zhu for their support with the 3D-FSI simulations. 
The authors would also like acknowledge the support from the Center for Research Computing at the University of Notre Dame for providing
computational resources that were essential to generate model results for this study.

\bibliographystyle{unsrt}
\bibliography{refs}{}

\begin{thebibliography}{10}

\bibitem{smith_our_2012}
Sidney~C. Smith, Amy Collins, Roberto Ferrari, David~R. Holmes, Susanne
  Logstrup, Diana~Vaca {McGhie}, Johanna Ralston, Ralph~L. Sacco, Hans Stam,
  Kathryn Taubert, David~A. Wood, and William~A. Zoghbi.
\newblock Our time: A call to save preventable death from cardiovascular
  disease (heart disease and stroke).
\newblock {\em Journal of the American College of Cardiology},
  60(22):2343--2348, 2012.

\bibitem{laslett_worldwide_2012}
Lawrence~J. Laslett, Peter Alagona, Bernard~A. Clark, Joseph~P. Drozda, Frances
  Saldivar, Sean~R. Wilson, Chris Poe, and Menolly Hart.
\newblock The worldwide environment of cardiovascular disease: Prevalence,
  diagnosis, therapy, and policy issues: A report from the american college of
  cardiology.
\newblock {\em Journal of the American College of Cardiology}, 60(25):S1--S49,
  2012.

\bibitem{QuarteroniAlfio2004MMaN}
Alfio Quarteroni and Luca Formaggia.
\newblock {\em Mathematical Modelling and Numerical Simulation of the
  Cardiovascular System}, volume~12 of {\em Computational Models for the Human
  Body}.
\newblock Elsevier, Amsterdam, 2004.

\bibitem{taylor_finite_1998}
Charles~A. Taylor, Thomas~J.R. Hughes, and Christopher~K. Zarins.
\newblock Finite element modeling of blood flow in arteries.
\newblock {\em Computer Methods in Applied Mechanics and Engineering},
  158(1):155--196, 1998.

\bibitem{liang_closed-loop_2005}
Fuyou Liang and Hao Liu.
\newblock A closed-loop lumped parameter computational model for human
  cardiovascular system.
\newblock {\em {JSME} International Journal Series C Mechanical Systems,
  Machine Elements and Manufacturing}, 48(4):484--493, 2005.

\bibitem{olufsen_dynamics_2002}
Mette~S. Olufsen, Ali Nadim, and Lewis~A. Lipsitz.
\newblock Dynamics of cerebral blood flow regulation explained using a lumped
  parameter model.
\newblock {\em American Journal of Physiology-Regulatory, Integrative and
  Comparative Physiology}, 282(2):R611--R622, 2002-02.
\newblock Publisher: American Physiological Society.

\bibitem{fossan_uncertainty_2018}
Fredrik~E. Fossan, Jacob Sturdy, Lucas~O. Müller, Andreas Strand, Anders~T.
  Bråten, Arve Jørgensen, Rune Wiseth, and Leif~R. Hellevik.
\newblock Uncertainty quantification and sensitivity analysis for computational
  {FFR} estimation in stable coronary artery disease.
\newblock {\em Cardiovascular Engineering and Technology}, 9(4):597--622, 2018.

\bibitem{zarins_computed_2013}
Christopher~K. Zarins, Charles~A. Taylor, and James~K. Min.
\newblock Computed fractional flow reserve ({FFTCT}) derived from coronary {CT}
  angiography.
\newblock {\em Journal of Cardiovascular Translational Research},
  6(5):708--714, 2013.

\bibitem{caroli_validation_2013}
Anna Caroli, Simone Manini, Luca Antiga, Katia Passera, Bogdan Ene-Iordache,
  Stefano Rota, Giuseppe Remuzzi, Aron Bode, Jaap Leermakers, Frans~N. van~de
  Vosse, Raymond Vanholder, Marko Malovrh, Jan Tordoir, Andrea Remuzzi, and {on
  behalf of the ARCH project Consortium}.
\newblock Validation of a patient-specific hemodynamic computational model for
  surgical planning of vascular access in hemodialysis patients.
\newblock {\em Kidney International}, 84(6):1237--1245, 2013.

\bibitem{bode_patient-specific_2012}
Aron~S. Bode, Wouter Huberts, E.~Marielle~H. Bosboom, Wilco Kroon, Wim P. M.
  van~der Linden, R.~Nils Planken, Frans N. van~de Vosse, and Jan H.~M.
  Tordoir.
\newblock Patient-specific computational modeling of upper extremity
  arteriovenous fistula creation: Its feasibility to support clinical
  decision-making.
\newblock {\em {PLOS} {ONE}}, 7(4):e34491, 2012.
\newblock Publisher: Public Library of Science.

\bibitem{steele_vivo_2001}
B.N. Steele, C.A. Taylor, J.~Wan, J.P. Ku, and T.J.R. Hughes.
\newblock In vivo validation of a one-dimensional finite element method for
  simulation-based medical planning for cardiovascular bypass surgery.
\newblock volume~1, pages 120--123 vol.1. 2001 Conference Proceedings of the
  23rd Annual International Conference of the {IEEE} Engineering in Medicine
  and Biology Society, 2001.
\newblock {ISSN}: 1094-687X.

\bibitem{fiusco_blood_2022}
Francesco Fiusco, Lars~Mikael Broman, and Lisa Prahl~Wittberg.
\newblock Blood pumps for extracorporeal membrane oxygenation: Platelet
  activation during different operating conditions.
\newblock {\em {ASAIO} journal (American Society for Artificial Internal
  Organs: 1992)}, 68(1):79--86, 2022.

\bibitem{fiusco_numerical_2023}
Francesco Fiusco, Federico Rorro, Lars~Mikael Broman, and Lisa Prahl~Wittberg.
\newblock Numerical and experimental investigation of a lighthouse tip drainage
  cannula used in extracorporeal membrane oxygenation.
\newblock {\em Artificial Organs}, 47(2):330--341, 2023.

\bibitem{migliavacca_expansion_2007}
F.~Migliavacca, F.~Gervaso, M.~Prosi, P.~Zunino, S.~Minisini, L.~Formaggia, and
  G.~Dubini.
\newblock Expansion and drug elution model of a coronary stent.
\newblock {\em Computer Methods in Biomechanics and Biomedical Engineering},
  10(1):63--73, 2007.
\newblock Publisher: Taylor \& Francis \_eprint:
  https://doi.org/10.1080/10255840601071087.

\bibitem{alderliesten_simulation_2004}
T~Alderliesten, Mk~Konings, and Wj~Niessen.
\newblock Simulation of minimally invasive vascular interventions for training
  purposes.
\newblock {\em Computer aided surgery : official journal of the International
  Society for Computer Aided Surgery}, 9(1), 2004.
\newblock Publisher: Comput Aided Surg.

\bibitem{stergiopulos_simple_1994}
N.~Stergiopulos, J.~J. Meister, and N.~Westerhof.
\newblock Simple and accurate way for estimating total and segmental arterial
  compliance: The pulse pressure method.
\newblock {\em Annals of Biomedical Engineering}, 22(4):392--397, 1994.

\bibitem{stergiopulos_total_1999}
Nikos Stergiopulos, Berend~E. Westerhof, and Nico Westerhof.
\newblock Total arterial inertance as the fourth element of the windkessel
  model.
\newblock {\em American Journal of Physiology-Heart and Circulatory
  Physiology}, 276(1):H81--H88, 1999.

\bibitem{anderson_verification_2007}
Andrew~E. Anderson, Benjamin~J. Ellis, and Jeffrey~A. Weiss.
\newblock Verification, validation and sensitivity studies in computational
  biomechanics.
\newblock {\em Computer Methods in Biomechanics and Biomedical Engineering},
  10(3):171--184, 2007.

\bibitem{food_and_drug_administration_assessing_2021}
{Food and Drug Administration}.
\newblock Assessing the credibility of computational modeling and simulation in
  medical device submissions, 2023.

\bibitem{saltelli_global_2008}
Andrea Saltelli, Matteo Ratto, Terry Andres, Francesca Campolongo, Jessica
  Cariboni, Debora Gatelli, Michaela Saisana, and Stefano Tarantola.
\newblock {\em Global sensitivity analysis the primer}.
\newblock John Wiley, 2008.

\bibitem{mcbook}
Art~B. Owen.
\newblock {\em Monte Carlo theory, methods and examples}.
\newblock \url{https://artowen.su.domains/mc/}, 2013.

\bibitem{xiu2002wiener}
Dongbin Xiu and George~Em Karniadakis.
\newblock The wiener--askey polynomial chaos for stochastic differential
  equations.
\newblock {\em SIAM journal on scientific computing}, 24(2):619--644, 2002.

\bibitem{sudret2008global}
Bruno Sudret.
\newblock Global sensitivity analysis using polynomial chaos expansions.
\newblock {\em Reliability engineering \& system safety}, 93(7):964--979, 2008.

\bibitem{eck_guide_2016}
Vinzenz~Gregor Eck, Wouter~Paulus Donders, Jacob Sturdy, Jonathan Feinberg,
  Tammo Delhaas, Leif~Rune Hellevik, and Wouter Huberts.
\newblock A guide to uncertainty quantification and sensitivity analysis for
  cardiovascular applications.
\newblock {\em International Journal for Numerical Methods in Biomedical
  Engineering}, 32(8):e02755, 2016.

\bibitem{tanade_global_2021}
Cyrus Tanade, Bradley Feiger, Madhurima Vardhan, S.~James Chen, Jane~A.
  Leopold, and Amanda Randles.
\newblock Global sensitivity analysis for clinically validated 1d models of
  fractional flow reserve.
\newblock In {\em 2021 43rd Annual International Conference of the {IEEE}
  Engineering in Medicine \& Biology Society ({EMBC})}, pages 4395--4398. EMBC,
  2021.
\newblock {ISSN}: 2694-0604.

\bibitem{eck_effects_2017}
V.G. Eck, J.~Sturdy, and L.R. Hellevik.
\newblock Effects of arterial wall models and measurement uncertainties on
  cardiovascular model predictions.
\newblock {\em Journal of Biomechanics}, 50:188--194, 2017.

\bibitem{zhang_personalized_2021}
Xiancheng Zhang, Jia Liu, Zaiheng Cheng, Bokai Wu, Jian Xie, Lin Zhang, Zhijun
  Zhang, and Hao Liu.
\newblock Personalized 0d-1d multiscale hemodynamic modeling and wave dynamics
  analysis of cerebral circulation for an elderly patient with dementia.
\newblock {\em International Journal for Numerical Methods in Biomedical
  Engineering}, 37(9):e3510, 2021.
\newblock \_eprint: https://onlinelibrary.wiley.com/doi/pdf/10.1002/cnm.3510.

\bibitem{gul_parametric_2015}
R.~Gul and S.~Bernhard.
\newblock Parametric uncertainty and global sensitivity analysis in a model of
  the carotid bifurcation: Identification and ranking of most sensitive model
  parameters.
\newblock {\em Mathematical Biosciences}, 269:104--116, 2015.

\bibitem{huberts_sensitivity_2013}
W.~Huberts, C.~De~Jonge, W.P.M. Van Der~Linden, M.A. Inda, K.~Passera, J.H.M.
  Tordoir, F.N. Van De~Vosse, and E.M.H. Bosboom.
\newblock A sensitivity analysis of a personalized pulse wave propagation model
  for arteriovenous fistula surgery. part b: Identification of possible generic
  model parameters.
\newblock {\em Medical Engineering \& Physics}, 35(6):827--837, 2013.

\bibitem{peherstorfer_optimal_2016}
Benjamin Peherstorfer, Karen Willcox, and Max Gunzburger.
\newblock Optimal model management for multifidelity monte carlo estimation.
\newblock {\em {SIAM} Journal on Scientific Computing}, 38(5):A3163--A3194,
  2016.

\bibitem{yin_one-dimensional_2019}
Minglang Yin, Alireza Yazdani, and George~Em Karniadakis.
\newblock One-dimensional modeling of fractional flow reserve in coronary
  artery disease: Uncertainty quantification and bayesian optimization.
\newblock {\em Computer Methods in Applied Mechanics and Engineering},
  353:66--85, 2019.

\bibitem{quaglino_fast_2018}
A.~Quaglino, S.~Pezzuto, P.~S. Koutsourelakis, A.~Auricchio, and R.~Krause.
\newblock Fast uncertainty quantification of activation sequences in
  patient-specific cardiac electrophysiology meeting clinical time constraints.
\newblock {\em International Journal for Numerical Methods in Biomedical
  Engineering}, 34(7):e2985, 2018.
\newblock \_eprint: https://onlinelibrary.wiley.com/doi/pdf/10.1002/cnm.2985.

\bibitem{seo_multifidelity_2020}
J.~Seo, C.~Fleeter, A.M. Kahn, A.L. Marsden, and D.E. Schiavazzi.
\newblock Multifidelity estimators for coronary circulation models under
  clinically informed data uncertainty.
\newblock {\em International Journal for Uncertainty Quantification},
  10(5):449--466, 2020.
\newblock Publisher: Begell House Inc.

\bibitem{fleeter_multilevel_2020}
Casey~M. Fleeter, Gianluca Geraci, Daniele~E. Schiavazzi, Andrew~M. Kahn, and
  Alison~L. Marsden.
\newblock Multilevel and multifidelity uncertainty quantification for
  cardiovascular hemodynamics.
\newblock {\em Computer Methods in Applied Mechanics and Engineering},
  365:113030, 2020.

\bibitem{qian_multifidelity_2018}
E.~Qian, B.~Peherstorfer, D.~O'Malley, V.~V. Vesselinov, and K.~Willcox.
\newblock Multifidelity monte carlo estimation of variance and sensitivity
  indices.
\newblock {\em {SIAM}/{ASA} Journal on Uncertainty Quantification},
  6(2):683--706, 2018.

\bibitem{du_uncertainty_2022}
Wenting Du and Jin Su.
\newblock Uncertainty quantification for numerical solutions of the nonlinear
  partial differential equations by using the multi-fidelity monte carlo
  method.
\newblock {\em Applied Sciences}, 12(14):7045, 2022.
\newblock Number: 14 Publisher: Multidisciplinary Digital Publishing Institute.

\bibitem{yao_multi-fidelity_2022}
Yuan Yao, Xun Huan, and Jesse Capecelatro.
\newblock Multi-fidelity uncertainty quantification of particle deposition in
  turbulent pipe flow.
\newblock {\em Journal of Aerosol Science}, 166:106065, 2022.

\bibitem{cataldo_multifidelity_2022}
Giuseppe Cataldo, Elizabeth Qian, and Jeremy Auclair.
\newblock Multifidelity uncertainty quantification and model validation of
  large-scale multidisciplinary systems.
\newblock {\em Journal of Astronomical Telescopes, Instruments, and Systems},
  8(3):038001, 2022.
\newblock Publisher: {SPIE}.

\bibitem{quaglino_high-dimensional_2019}
A.~Quaglino, S.~Pezzuto, and R.~Krause.
\newblock High-dimensional and higher-order multifidelity monte carlo
  estimators.
\newblock {\em Journal of Computational Physics}, 388:300--315, 2019.

\bibitem{pagani_enabling_2021}
Stefano Pagani and Andrea Manzoni.
\newblock Enabling forward uncertainty quantification and sensitivity analysis
  in cardiac electrophysiology by reduced order modeling and machine learning.
\newblock {\em International Journal for Numerical Methods in Biomedical
  Engineering}, 37(6):e3450, 2021.
\newblock \_eprint: https://onlinelibrary.wiley.com/doi/pdf/10.1002/cnm.3450.

\bibitem{saltelli_variance_2010}
Andrea Saltelli, Paola Annoni, Ivano Azzini, Francesca Campolongo, Marco Ratto,
  and Stefano Tarantola.
\newblock Variance based sensitivity analysis of model output. design and
  estimator for the total sensitivity index.
\newblock {\em Computer Physics Communications}, 181(2):259--270, 2010.

\bibitem{owen_variance_2013}
Art~B. Owen.
\newblock Variance components and generalized sobol' indices.
\newblock {\em {SIAM}/{ASA} Journal on Uncertainty Quantification},
  1(1):19--41, 2013.

\bibitem{sudret_global_2008}
Bruno Sudret.
\newblock Global sensitivity analysis using polynomial chaos expansions.
\newblock 93(7):964--979.

\bibitem{bazilevs_isogeometric_2008}
Y.~Bazilevs, V.~M. Calo, T.~J.~R. Hughes, and Y.~Zhang.
\newblock Isogeometric fluid-structure interaction: theory, algorithms, and
  computations.
\newblock {\em Computational Mechanics}, 43(1):3--37, 2008.

\bibitem{geuzaine_gmsh_2009}
Christophe Geuzaine and Jean-François Remacle.
\newblock Gmsh: A 3-d finite element mesh generator with built-in pre- and
  post-processing facilities.
\newblock {\em International Journal for Numerical Methods in Engineering},
  79(11):1309--1331, 2009.
\newblock \_eprint: https://onlinelibrary.wiley.com/doi/pdf/10.1002/nme.2579.

\bibitem{zhu_svfsi_2022}
Chi Zhu, Vijay Vedula, Dave Parker, Nathan Wilson, Shawn Shadden, and Alison
  Marsden.
\newblock {svFSI}: A multiphysics package for integrated cardiac modeling.
\newblock {\em Journal of Open Source Software}, 7(78):4118, 2022.

\bibitem{updegrove_simvascular_2017}
Adam Updegrove, Nathan~M. Wilson, Jameson Merkow, Hongzhi Lan, Alison~L.
  Marsden, and Shawn~C. Shadden.
\newblock {SimVascular}: An open source pipeline for cardiovascular simulation.
\newblock {\em Annals of Biomedical Engineering}, 45(3):525--541, 2017.

\bibitem{lan_re-engineered_2018}
Hongzhi Lan, Adam Updegrove, Nathan~M. Wilson, Gabriel~D. Maher, Shawn~C.
  Shadden, and Alison~L. Marsden.
\newblock A re-engineered software interface and workflow for the open-source
  {SimVascular} cardiovascular modeling package.
\newblock {\em Journal of Biomechanical Engineering}, 140(2):0245011--02450111,
  2018.

\bibitem{heroux_overview_2005}
Michael~A. Heroux, Roscoe~A. Bartlett, Vicki~E. Howle, Robert~J. Hoekstra,
  Jonathan~J. Hu, Tamara~G. Kolda, Richard~B. Lehoucq, Kevin~R. Long, Roger~P.
  Pawlowski, Eric~T. Phipps, Andrew~G. Salinger, Heidi~K. Thornquist, Ray~S.
  Tuminaro, James~M. Willenbring, Alan Williams, and Kendall~S. Stanley.
\newblock An overview of the trilinos project.
\newblock {\em {ACM} Transactions on Mathematical Software}, 31(3):397--423,
  2005.

\bibitem{seo_performance_2019}
Jongmin Seo, Daniele~E. Schiavazzi, and Alison~L. Marsden.
\newblock Performance of preconditioned iterative linear solvers for
  cardiovascular simulations in rigid and deformable vessels.
\newblock {\em Computational mechanics}, 64:717--739, 2019.

\bibitem{hsu_blood_2011}
Ming-Chen Hsu and Yuri Bazilevs.
\newblock Blood vessel tissue prestress modeling for vascular fluid–structure
  interaction simulation.
\newblock {\em Finite Elements in Analysis and Design}, 47(6):593--599, 2011.

\bibitem{jansen2000generalized}
Kenneth~E Jansen, Christian~H Whiting, and Gregory~M Hulbert.
\newblock A generalized-$\alpha$ method for integrating the filtered
  {N}avier--{S}tokes equations with a stabilized finite element method.
\newblock {\em Computer methods in applied mechanics and engineering},
  190(3-4):305--319, 2000.

\bibitem{sherwin_one-dimensional_2003}
S.J. Sherwin, V.~Franke, J.~Peiró, and K.~Parker.
\newblock One-dimensional modelling of a vascular network in space-time
  variables.
\newblock {\em Journal of Engineering Mathematics}, 47(3):217--250, 2003.

\bibitem{boileau_benchmark_2015}
Etienne Boileau, Perumal Nithiarasu, Pablo~J. Blanco, Lucas~O. Müller,
  Fredrik~Eikeland Fossan, Leif~Rune Hellevik, Wouter~P. Donders, Wouter
  Huberts, Marie Willemet, and Jordi Alastruey.
\newblock A benchmark study of numerical schemes for one-dimensional arterial
  blood flow modelling.
\newblock {\em International Journal for Numerical Methods in Biomedical
  Engineering}, 31(10):e02732, 2015.

\bibitem{eck_uncertainty_2016}
Vinzenz~Gregor Eck.
\newblock Uncertainty quantification and sensitivity analysis for
  cardiovascular models.
\newblock Doctoral thesis, NTNU.
\newblock Accepted: 2016-10-11T13:39:25Z {ISBN}: 9788232618071 {ISSN}:
  1503-8181.

\bibitem{figueroa_coupled_2006}
C.~Alberto Figueroa, Irene~E. Vignon-Clementel, Kenneth~E. Jansen, Thomas~J.R.
  Hughes, and Charles~A. Taylor.
\newblock A coupled momentum method for modeling blood flow in
  three-dimensional deformable arteries.
\newblock {\em Computer Methods in Applied Mechanics and Engineering},
  195(41):5685--5706, 2006.

\bibitem{xiao_systematic_2014}
Nan Xiao, Jordi Alastruey, and C.~Alberto Figueroa.
\newblock A systematic comparison between 1-d and 3-d hemodynamics in compliant
  arterial models.
\newblock {\em International Journal for Numerical Methods in Biomedical
  Engineering}, 30(2):204--231, 2014.
\newblock \_eprint: https://onlinelibrary.wiley.com/doi/pdf/10.1002/cnm.2598.

\bibitem{pfaller_automated_2022}
Martin~R. Pfaller, Jonathan Pham, Aekaansh Verma, Luca Pegolotti, Nathan~M.
  Wilson, David~W. Parker, Weiguang Yang, and Alison~L. Marsden.
\newblock Automated generation of 0d and 1d reduced-order models of
  patient-specific blood flow.
\newblock {\em International Journal for Numerical Methods in Biomedical
  Engineering}, 38(10):e3639, 2022.
\newblock \_eprint: https://onlinelibrary.wiley.com/doi/pdf/10.1002/cnm.3639.

\bibitem{vriz_comparison_2017}
Olga Vriz, Julien Magne, Caterina Driussi, Gabriele Brosolo, Francesco Ferrara,
  Paolo Palatini, Victor Aboyans, and Eduardo Bossone.
\newblock Comparison of arterial stiffness/compliance in the ascending aorta
  and common carotid artery in healthy subjects and its impact on left
  ventricular structure and function.
\newblock {\em The International Journal of Cardiovascular Imaging},
  33(4):521--531, 2017.

\bibitem{hansen_diameter_1995}
F.~Hansen, P.~Mangell, B.~Sonesson, and T.~Länne.
\newblock Diameter and compliance in the human common carotid artery —
  variations with age and sex.
\newblock {\em Ultrasound in Medicine \& Biology}, 21(1):1--9, 1995.

\bibitem{raninen_carotid_2002}
Reino~O. Raninen, Markku~M. Kupari, and Pauli~E. Hekali.
\newblock Carotid and femoral artery stiffnessin takayasu's arteritis.
\newblock {\em Scandinavian Journal of Rheumatology}, 31(2):85--88, 2002.

\bibitem{riley_variation_1997}
Ward~A. Riley, Gregory~W. Evans, A.~Richey~Sharrett, Gregory~L. Burke, and
  Ralph~W. Barnes.
\newblock Variation of common carotid artery elasticity with intimal-medial
  thickness: The aric study.
\newblock {\em Ultrasound in Medicine \& Biology}, 23(2):157--164, 1997.

\bibitem{engelen_reference_2015}
Lian Engelen, Jelle Bossuyt, Isabel Ferreira, Luc~M. van Bortel, Koen~D.
  Reesink, Patrick Segers, Coen~D. Stehouwer, Stéphane Laurent, and Pierre
  Boutouyrie.
\newblock Reference values for local arterial stiffness. part a: carotid
  artery.
\newblock {\em Journal of Hypertension}, 33(10):1981--1996, 2015.

\bibitem{schiavazzi_multifidelity_2020}
D.E. Schiavazzi, C.M. Fleeter, G.~Geraci, and A.L. Marsden.
\newblock Multifidelity uncertainty propagation for cardiovascular
  hemodynamics.
\newblock pages 2759--2770. Proc. Eur. Conf. Comput. Mech.: Solids, Struct.
  Coupled Probl., {ECCM} Eur. Conf. Comput. Fluid Dyn., {ECFD}, 2020.

\bibitem{ceserani_lumped-parameter_2023}
Valentina Ceserani, Mauro Lo~Rito, Mauro~Luca Agnifili, Ariel~F. Pascaner,
  Antonio Rosato, Serena Anglese, Miriam Deamici, Jessica Negri, Chiara
  Corrado, Francesco Bedogni, Francesco Secchi, Massimo Lombardi, Ferdinando
  Auricchio, Alessandro Frigiola, and Michele Conti.
\newblock Lumped-parameter model as a non-invasive tool to assess coronary
  blood flow in {AAOCA} patients.
\newblock {\em Scientific Reports}, 13(1):17448, 2023.
\newblock Number: 1 Publisher: Nature Publishing Group.

\bibitem{mantero_coronary_1992}
S.~Mantero, R.~Pietrabissa, and R.~Fumero.
\newblock The coronary bed and its role in the cardiovascular system: a review
  and an introductory single-branch model.
\newblock {\em Journal of Biomedical Engineering}, 14(2):109--116, 1992.

\bibitem{shi_review_2011}
Yubing Shi, Patricia Lawford, and Rodney Hose.
\newblock Review of zero-d and 1-d models of blood flow in the cardiovascular
  system.
\newblock {\em {BioMedical} Engineering {OnLine}}, 10(1):33, 2011.

\end{thebibliography}

\subsection*{Appendix}

\begin{landscape}
    \footnotesize
    \begin{table}[]
    \centering
    \caption{Model allocations for 1D/0D bi-fidelity SA for all QoIs at increasing computational budgets. $m_k$ represents the model evaluations of the set $\{y^{(A,s)} \}_{s=1}^N$, while the total number of model evaluations is the sum of $\{y^{(A,s)} \}_{s=1}^N$, $\{y^{(B,s)} \}_{s=1}^N$, and $\{y^{(C_j,s)} \}_{s=1}^N$, thus $(d+2)\,m_k = 5\,m_k$. The perturbed and unperturbed 0D-model statistics led to the same model allocations.}
    \begin{tabular}{c|l|c |c|c c |c c |c c| c c| c c| c c| c c}
        \toprule
         & & & & \multicolumn{14}{c}{computational budget} \\
         & QoI & model & $\alpha_k$ & \multicolumn{2}{c}{500} & \multicolumn{2}{c}{1000} & \multicolumn{2}{c}{2000} & \multicolumn{2}{c}{4000} & \multicolumn{2}{c}{6000} & \multicolumn{2}{c}{8000} & \multicolumn{2}{c}{10000} \\
         & & & & $m_k$ & total & $m_k$ & total & $m_k$ & total & $m_k$ & total & $m_k$ & total & $m_k$ & total & $m_k$ & total\\
         \midrule
        \multirow{6}{*}{\rotatebox[origin=c]{90}{unperturbed}} & \multirow{2}{*}{P$_{\text{sys}}$} & 1D & 1 & 4 & 20 & 9 & 45 & 18 & 90 & 37 & 185 & 55 & 275 & 74 & 370 & 92 & 460\\
         & & 0D & 0.9916 & 317 & 1585 & 635 & 3175 & 1271 & 6355 & 2542 & 12710 & 3814 & 19710 & 5085 & 25425 & 6356 & 31780\\
         \cline{2-18}
        & \multirow{2}{*}{PP} & 1D & 1 & 3 & 15 & 6 & 30 & 12 & 60 & 24 & 120 & 37 & 185 & 49 & 245 & 61 & 305\\
         & & 0D & 1.0114 & 323 & 1615 & 646 & 3230 & 1292 & 6460 & 2584 & 12920 & 3876 & 19380 & 5168 & 25840 & 6460 & 32300 \\
         \cline{2-18}
        & \multirow{2}{*}{$\Delta r_{\text{max}}$} & 1D & 1 & 2 & 10 & 3 & 15 & 6 & 30 & 13 & 65 & 19 & 95 & 26 & 130 & 33 & 165 \\
         & & 0D & 0.9598 & 327 & 1635 & 655 & 3275 & 1311 & 6555 & 2622 & 13110 & 3933 & 19665 & 5244 & 26220 & 6555 & 32775 \\
                  \midrule

        \multirow{6}{*}{\rotatebox[origin=c]{90}{perturbed}} & \multirow{2}{*}{P$_{\text{sys}}$} & 1D & 1 & 21 & 105 & 43 & 215 & 87 & 435 & 175 & 875 & 262 & 1310  & 350 & 1750 & 437 & 2185\\
         & & 0D & 0.8611 & 260 & 1300 & 520 & 2600 & 1041 & 5205 & 2083 & 10415 & 3124 & 15620 & 4166 & 20830 & 5208 & 26040\\
         \cmidrule{2-18}
        & \multirow{2}{*}{PP} & 1D & 1 & 16 & 80 & 32 & 160& 65 & 325 & 130 & 650 & 196 & 980 & 261 & 1305 & 326 & 1630\\
         & & 0D & 0.9769 &  278 & 1390 & 520 & 2600 & 1115 & 5575 & 2230 & 11150 & 3346 & 16730 & 4461 & 22305 & 5576 & 27880 \\
         \cmidrule{2-18}
        & \multirow{2}{*}{$\Delta r_{\text{max}}$} & 1D & 1 & 7 & 35 & 14 & 70 & 29 & 145 & 59 & 295 & 88 & 440 & 118 & 590 & 148 & 740 \\
         & & 0D & 0.9105 & 308 & 1540  & 617 & 3085 & 1234 & 6170 & 2468 & 12340 & 3703 & 18515 & 4937 & 24685 & 6172 & 30860\\
         \bottomrule
    \end{tabular}
    \label{tab:ModelAllocation_1D_0D_extended}
\end{table}
\end{landscape}

\begin{landscape}
    \begin{table}[]
    \centering
        \caption{Model response statistics for tri-fidelity estimator from 10, 20, or 150 pilot runs. The statistics are the mean~$\mu$, the standard deviation~$\sigma$, the correlation with the 1D-model~$\rho_{1D,dD}$, higher order statistics $\delta_k$, $\tau$, q, and the computational costs $w$. The maximal difference~$\Delta$ in estimated model response statistics of any QoI between 10 and 150 or 20 and 150 pilot runs is given in the last column.} 
        \footnotesize
    \resizebox{1.4\textwidth}{!}{        
    \begin{tabular}{l c |c c c c c|c c c c c|c c c c c | c}
        \toprule
        \# &\textbf{}  & \multicolumn{5}{c}{P$_{sys}$} &\multicolumn{5}{c}{PP}  & \multicolumn{5}{c}{$\Delta r_{\text{max}}$} \\ 
        & & 3D & 1D & 0D & 0D pert. & unit & 3D & 1D & 0D & 0D pert. & unit& 3D & 1D & 0D & 0D pert. & unit & max. $\Delta$ [\%]\\
        \midrule

        \multirow{6}{*}{10} & $\mu$ &  129.4 & 129.6 & 130.7 & 130.7 & mmHg & 50.4 & 50.9 & 51.7 & 51.7 & mmHg & 0.173 & 0.163 & 0.164 & 0.163 & mm & 0.6\\
        & $\sigma$ &  1.68 & 1.56 & 1.59 & 1.45 & mmHg & 2.53 & 2.41 & 2.41 & 2.27 & mmHg & 0.015 & 0.013 & 0.013 & 0.011 & mm & 21.5 \\
        & $\rho_{3D,dD}$ &  1 & 0.9826 & 0.9827 & 0.9650 & - &  1 & 0.9809 & 0.9811 & 0.9619 & - & 1 & 0.9495 & 0.9490 & 0.8382 & - & 5.6 \\
        & $\delta_k$ &  30.38 & 25.23 & 26.50 & 20.06 & mmHg$^4$ & 152.22 & 142.28 & 140.57 & 119.90 & mmHg$^4$ & 2.30$\cdot 10^{-7}$ & 1.46$\cdot 10^{-7}$ & 1.53$\cdot 10^{-7}$ & 8.98$\cdot 10^{-8}$ & mm$^4$ & 46.0 \\
        & $\tau_k$ &  527.9 & 487.7 & 498.3 & 498.3 & mmHg & 1176.2 & 1156.4 & 1147.4 & 1147.4 & mmHg & 3.50$\cdot 10^{-7}$ & 2.87$\cdot 10^{-7}$ & 2.94$\cdot 10^{-7}$ & 2.94$\cdot 10^{-7}$ & mm & 23.5 \\
        & $q_{3D,dD}$ &  1 & 0.9938 & 0.9935 & 0.9837 & - & 1 & 0.9932 & 0.9930 & 0.9838 & - & 1 & 0.9747 & 0.9742 & 0.9319 & - & 18.1 \\
         \midrule
         
        \multirow{6}{*}{20} & $\mu$ &  129.6 & 129.7 & 130.8 & 130.8 & mmHg & 50.6 & 51.1 & 51.9 & 51.9 & mmHg & 0.172 & 0.162 & 0.163 & 0.163 & mm & 0.1 \\
        & $\sigma$ &  1.98 & 1.84 & 1.87 & 1.78 & mmHg & 3.02 & 2.86 & 2.85 & 2.71 & mmHg & 0.017 & 0.014 & 0.015 & 0.013 & mm & 7.1 \\
        & $\rho_{3D,dD}$ &  1 & 0.9915 & 0.9919 & 0.9878 & - & 1 & 0.9917 & 0.9921 & 0.9903 & - & 1 & 0.9678 & 0.9673 & 0.9619 & - & 8.3 \\
        & $\delta_k$ &  32.86 & 27.46 & 28.91 & 23.02 & mmHg$^4$ & 188.28 & 163.33 & 160.90 & 129.78 & mmHg$^4$ & 1.94$\cdot 10^{-7}$ & 1.15$\cdot 10^{-7}$ & 1.22$\cdot 10^{-7}$ & 7.09$\cdot 10^{-8}$ & mm$^4$ & 24.0 \\
        & $\tau_k$ &  524.0 & 498.8 & 509.7 & 509.7 & mmHg & 1277.1 & 1221.6 & 1209.1 & 1209.1 & mmHg & 3.10$\cdot 10^{-7}$ & 2.47$\cdot 10^{-7}$ & 2.55$\cdot 10^{-7}$ & 2.55$\cdot 10^{-7}$ & mm & 7.1 \\
        & $q_{3D,dD}$ &  1 & 0.986 & 0.9874 & 0.986 & - & 1 & 0.9909 & 0.9915 & 0.9916 & - & 1 & 0.9400 & 0.9399 & 0.9381 & -  & 18.9\\
         
         \midrule         
          \multirow{6}{*}{150} & $\mu$ &  129.5 & 129.7 & 130.8 & 130.8 & mmHg & 50.6 & 51.1 & 51.9 & 51.9 & mmHg & 0.172 & 0.162 & 0.163 & 0.163 & mm & - \\
          & $\sigma$ &  1.93 & 1.80 & 1.84 & 1.79 & mmHg & 2.94 & 2.80 & 2.80 & 2.75 & mmHg & 0.016 & 0.014 & 0.014 & 0.014 & mm & - \\
          & $\rho_{3D,dD}$ & 1 & 0.9813 & 0.9835 & 0.969 & - & 1 & 0.9802 & 0.9817 & 0.9643 & - & 1 & 0.9422 & 0.9412 & 0.8879 & - & - \\
          & $\delta_k$ &  28.96 & 23.42 & 24.84 & 23.70 & mmHg$^4$& 155.97 & 138.65 & 137.19 & 134.83 & mmHg$^4$& 1.66$\cdot 10^{-7}$ & 1.00$\cdot 10^{-7}$ & 1.08$\cdot 10^{-7}$ & 9.33$\cdot 10^{-8}$ & mm$^4$ & - \\
        & $\tau_k$ &  512.1 & 474.0 & 484.8 &  483.2 & mmHg & 1193.2 & 1159.1 & 1147.5 & 1163.3 & mmHg & 3.06$\cdot 10^{-7}$ & 2.44$\cdot 10^{-7}$ & 2.53$\cdot 10^{-7}$ & 2.38$\cdot 10^{-7}$ & mm & - \\
        & $q_{3D,dD}$ & 1.0 & 0.9408 & 0.9459 & 0.9132 & - & 1.0 & 0.9402 & 0.9434 & 0.9075 & - & 1.0 & 0.8536 & 0.8527 & 0.7890 & - & - \\
         \midrule
         & $w$ & 1 & $9\cdot 10^{-5}$ & $3\cdot 10^{-5}$ & $3\cdot 10^{-5}$ & -& 1 & $9\cdot 10^{-5}$ & $3\cdot 10^{-5}$ & $3\cdot 10^{-5}$ & - & 1 & $9\cdot 10^{-5}$ & $3\cdot 10^{-5}$ & $3\cdot 10^{-5}$ & - & - \\
        
         \bottomrule
    \end{tabular}}
    \label{tab:ModelStatistics_3D_1D_0D_extended}
\end{table} 
\end{landscape}

\begin{figure}[!ht]
    \centering
    \includegraphics[width=\textwidth]{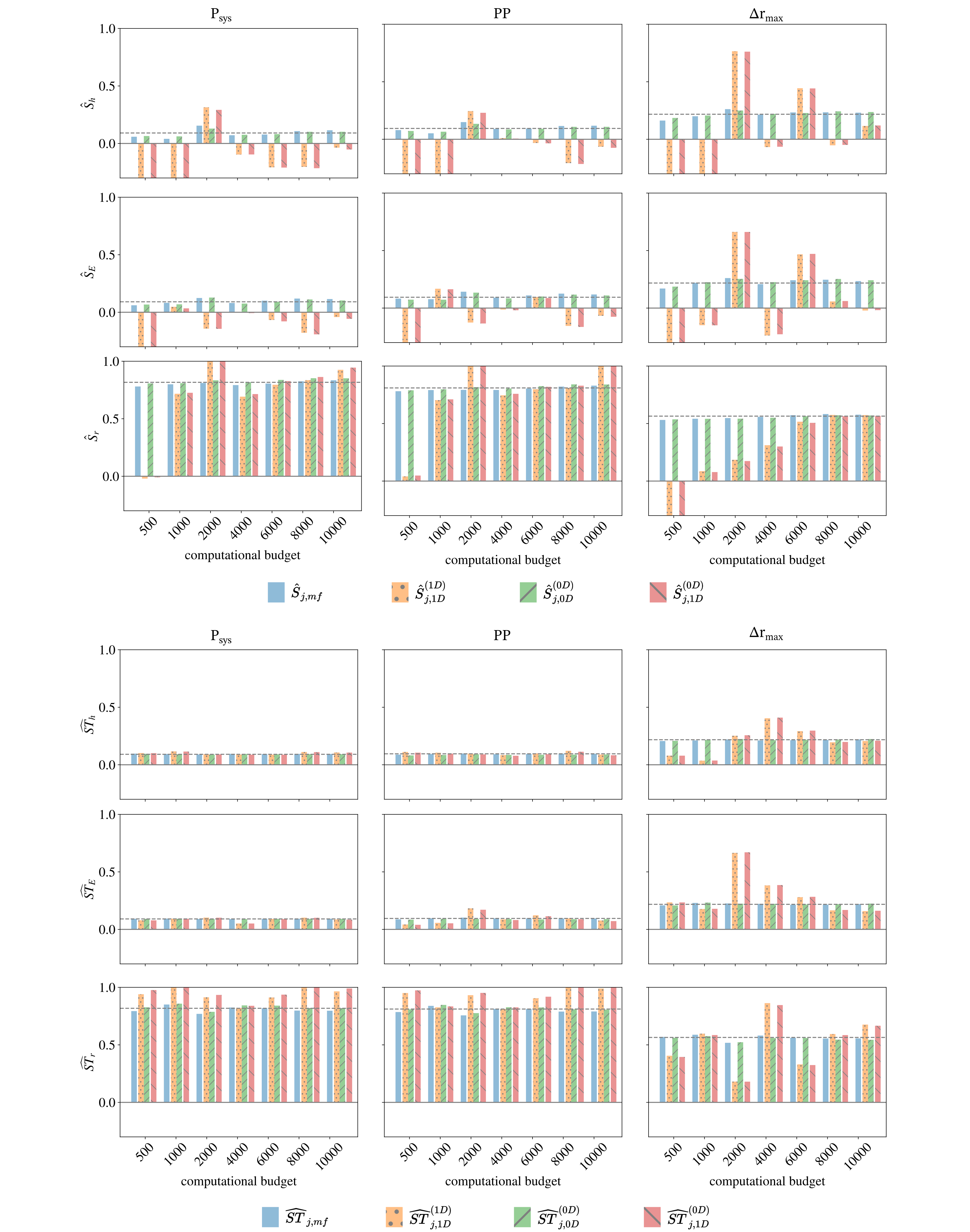}
    \caption{The upper panels show the unperturbed bi-fidelity estimates for the main Sobol' sensitivity indices $\hat{S}_{j,mf}$ for the QoIs P$_{\mathrm{sys}}$, PP, and $\Delta r_{\mathrm{max}}$ for computational budgets 500-10000, as well as the intermediate sensitivity results of $\hat{S}_{j,1D}^{1D}$, $\hat{S}_{j,0D}^{0D}$, and $\hat{S}_{j,1D}^{0D}$. 
    The lower panels are the analogs for the total sensitivity indices $\widehat{ST}_{j,mf}$, $\widehat{ST}_{j,1D}^{1D}$, $\widehat{ST}_{j,0D}^{0D}$, and $\widehat{ST}_{j,1D}^{0D}$. For comparison, fourth order polynomial chaos estimates of the same sensitivity indices are shown with horizontal dashed lines.}
    \label{fig:Intermediate_1D_0D}
\end{figure} 

\begin{table}[!ht]
    \scriptsize
    \centering
    \caption{Mean values and standard deviations of the main and total sensitivity indices for an increasing budget, computed using 100 realizations of the bi-fidelity estimator.}
    \label{tab:MFMC_Variations_S_ST}
    \begin{tabular}{l|c|cc|cc|cc|cc|cc|cc}
    \toprule
    \multirow{2}{*}{QoI} & \multirow{2}{*}{budget} & \multicolumn{2}{c}{$\hat{S}_h$} & \multicolumn{2}{c}{$\hat{S}_E$} & \multicolumn{2}{c}{$\hat{S}_r$} & \multicolumn{2}{c}{$\widehat{ST}_h$} & \multicolumn{2}{c}{$\widehat{ST}_E$} & \multicolumn{2}{c}{$\widehat{ST}_r$}\\
    & & $\mu$ & $\sigma$ & $\mu$ & $\sigma$ & $\mu$ & $\sigma$ & $\mu$ & $\sigma$ & $\mu$ & $\sigma$ & $\mu$ & $\sigma$ \\
\midrule
\multirow{4}{*}{P$_{sys}$} & 500 & 0.085 & 0.060 & 0.085 & 0.062 & 0.825 & 0.056 & 0.099 & 0.019 & 0.095 & 0.014 & 0.839 & 0.064 \\
 & 1000 & 0.094 & 0.039 & 0.091 & 0.04 & 0.814 & 0.032 & 0.095 & 0.010 & 0.093 & 0.007 & 0.819 & 0.040 \\
 & 2000 & 0.084 & 0.027 & 0.083 & 0.025 & 0.815 & 0.027 & 0.093 & 0.004 & 0.093 & 0.006 & 0.826 & 0.025 \\
 & 4000 & 0.092 & 0.022 & 0.092 & 0.023 & 0.813 & 0.017 & 0.093 & 0.004 & 0.093 & 0.004 & 0.816 & 0.019 \\
\midrule
\multirow{4}{*}{PP} & 500 & 0.09 & 0.061 & 0.088 & 0.057 & 0.816 & 0.056 & 0.100 & 0.021 & 0.098 & 0.014 & 0.831 & 0.075 \\
& 1000 & 0.095 & 0.038 & 0.093 & 0.039 & 0.807 & 0.029 & 0.098 & 0.009 & 0.097 & 0.007 & 0.814 & 0.040 \\
& 2000 & 0.087 & 0.026 & 0.085 & 0.025 & 0.808 & 0.026 & 0.096 & 0.004 & 0.096 & 0.005 & 0.802 & 0.025 \\
& 4000 & 0.094 & 0.022 & 0.095 & 0.023 & 0.807 & 0.016 & 0.095 & 0.003 & 0.096 & 0.003 & 0.811 & 0.020 \\
\midrule
\multirow{4}{*}{$\Delta r_{\text{max}}$} & 500 & 0.213 & 0.063 & 0.209 & 0.067 & 0.581 & 0.072 & 0.225 & 0.072 & 0.223 & 0.046 & 0.59 & 0.071 \\
& 1000 & 0.214 & 0.044 & 0.213 & 0.043 & 0.561 & 0.042 & 0.218 & 0.027 & 0.222 & 0.022 & 0.572 & 0.042 \\
& 2000 & 0.211 & 0.028 & 0.209 & 0.025 & 0.564 & 0.024 & 0.218 & 0.007 & 0.217 & 0.010 & 0.575 & 0.023 \\
& 4000 & 0.218 & 0.022 & 0.218 & 0.023 & 0.565 & 0.017 & 0.217 & 0.006 & 0.218 & 0.006 & 0.565 & 0.016 \\
\bottomrule
\end{tabular}
\end{table}

\begin{table}[!ht]
    \scriptsize
    \centering
    \caption{Mean values and standard deviations of the main and total sensitivity indices over 100 repetitions of the 1D model estimated with the Monte Carlo method.}
    \label{tab:MC_Variations_S_ST}
    \begin{tabular}{l|c|cc|cc|cc|cc|cc|cc}
    \toprule
    \multirow{2}{*}{QoI} & \multirow{2}{*}{budget} & \multicolumn{2}{c}{$\hat{S}_h$} & \multicolumn{2}{c}{$\hat{S}_E$} & \multicolumn{2}{c}{$\hat{S}_r$} & \multicolumn{2}{c}{$\widehat{ST}_h$} & \multicolumn{2}{c}{$\widehat{ST}_E$} & \multicolumn{2}{c}{$\widehat{ST}_r$}\\
    & & $\mu$ & $\sigma$ & $\mu$ & $\sigma$ & $\mu$ & $\sigma$ & $\mu$ & $\sigma$ & $\mu$ & $\sigma$ & $\mu$ & $\sigma$ \\
\midrule
\multirow{4}{*}{P$_{sys}$} & 500 & 0.093 & 0.113 & 0.089 & 0.115 & 0.814 & 0.088 & 0.093 & 0.013 & 0.092 & 0.014 & 0.828 & 0.098 \\
 & 1000 & 0.097 & 0.064 & 0.089 & 0.066 & 0.812 & 0.051 & 0.093 & 0.009 & 0.093 & 0.009 & 0.810 & 0.063 \\
 & 2000 & 0.085 & 0.053 & 0.087 & 0.049 & 0.816 & 0.042 & 0.093 & 0.007 & 0.093 & 0.007 & 0.828 & 0.050 \\
 & 4000 & 0.093 & 0.037 & 0.090 & 0.035 & 0.820 & 0.029 & 0.094 & 0.005 & 0.093 & 0.004 & 0.821 & 0.035 \\
\midrule
 \multirow{4}{*}{PP} & 500 & 0.096 & 0.113 & 0.092 & 0.115 & 0.808 & 0.088 & 0.097 & 0.014 & 0.095 & 0.014 & 0.822 & 0.098 \\
 & 1000 & 0.100 & 0.064 & 0.092 & 0.067 & 0.805 & 0.052 & 0.096 & 0.009 & 0.096 & 0.010 & 0.804 & 0.063 \\
 & 2000 & 0.088 & 0.053 & 0.089 & 0.049 & 0.810 & 0.042 & 0.096 & 0.007 & 0.096 & 0.007 & 0.823 & 0.050 \\
 & 4000 & 0.096 & 0.037 & 0.093 & 0.035 & 0.813 & 0.029 & 0.097 & 0.005 & 0.096 & 0.005 & 0.816 & 0.035 \\
 \midrule
 \multirow{4}{*}{$\Delta r_{\text{max}}$} & 500 & 0.224 & 0.109 & 0.220 & 0.115 & 0.567 & 0.094 & 0.220 & 0.033 & 0.217 & 0.033 & 0.574 & 0.074 \\
 & 1000 & 0.221 & 0.062 & 0.211 & 0.067 & 0.563 & 0.061 & 0.219 & 0.021 & 0.218 & 0.022 & 0.561 & 0.050 \\
 & 2000 & 0.215 & 0.052 & 0.216 & 0.046 & 0.566 & 0.048 & 0.219 & 0.016 & 0.219 & 0.016 & 0.575 & 0.039 \\
 & 4000 & 0.221 & 0.034 & 0.216 & 0.036 & 0.568 & 0.032 & 0.220 & 0.012 & 0.219 & 0.010 & 0.569 & 0.028 \\
\bottomrule
\end{tabular}
\end{table}

\begin{table}[!ht]
    \scriptsize
    \centering
        \caption{Mean values and standard deviations of the main and total sensitivity indices over 6~repetitions estimated with the tri-fidelity MFMC method.}
    \begin{tabular}{l|cc|cc|cc|cc|cc|cc}
    \toprule
    \multirow{2}{*}{QoI}  & \multicolumn{2}{c}{$\hat{S}_h$} & \multicolumn{2}{c}{$\hat{S}_E$} & \multicolumn{2}{c}{$\hat{S}_r$} & \multicolumn{2}{c}{$\widehat{ST}_h$} & \multicolumn{2}{c}{$\widehat{ST}_E$} & \multicolumn{2}{c}{$\widehat{ST}_r$}\\
    & $\mu$ & $\sigma$ & $\mu$ & $\sigma$ & $\mu$ & $\sigma$ & $\mu$ & $\sigma$ & $\mu$ & $\sigma$ & $\mu$ & $\sigma$ \\
        \midrule
        P$_{\mathrm{sys}}$  & 0.030 & 0.075 & 0.036 & 0.067 & 0.820 & 0.048 & 0.084 & 0.026 & 0.080 & 0.014 & 0.883 & 0.068 \\
        PP  & 0.063 & 0.074 & 0.053 & 0.055 & 0.824 & 0.046 & 0.087 & 0.028 & 0.077 & 0.017 & 0.859 & 0.064 \\
        $\Delta r_{\mathrm{max}}$  & 0.185 & 0.314 & 0.067 & 0.172 & 0.81 & 0.195 & 0.244 & 0.319 & 0.195 & 0.112 & 0.994 & 0.742 \\
        \bottomrule
    \end{tabular}
    \label{tab:triFidelity_Variations_S_ST}
\end{table}

\end{document}